\documentclass[%
 reprint,
floatfix,
nofootinbib,
 amsmath,amssymb,
 aps,
 pra,
]{revtex4-2}
\usepackage{mathtools}
\usepackage{graphicx}% Include figure files
\usepackage{bm}% bold math
\usepackage{hyperref}% add hypertext capabilities
\usepackage[T1]{fontenc}
\usepackage{color}
\usepackage{multirow}

\newcommand*{\Zero}{\textbf{0}}

\newcommand{\blockdiagonal}[3]{
\begin{bmatrix}
    #1      & \Zero     & \cdots    & \Zero     \\
    \Zero   & #2        & \cdots    & \Zero     \\
    \vdots  & \vdots    & \ddots    & \vdots    \\
    \Zero   & \Zero     & \cdots    & #3        \\
\end{bmatrix}
}

\newtheorem{exa}{Example}

\begin{document}

\date{\today}
\title{Quantum Circuit Overhead}

\author{Oskar S{\l}owik }
\email{oslowik@cft.edu.pl}
\affiliation{Center for Theoretical Physics, Polish Academy of Sciences,\\
Aleja Lotnik\'ow 32/46, 02-668 Warszawa, Poland}
\author{Piotr Dulian}
\affiliation{Center for Theoretical Physics, Polish Academy of Sciences,\\
Aleja Lotnik\'ow 32/46, 02-668 Warszawa, Poland}
\affiliation{Centre for Quantum Optical Technologies, Centre of New Technologies, University of Warsaw, Banacha 2c, 02-097 Warsaw, Poland}
\author{Adam Sawicki}
\email{a.sawicki@cft.edu.pl}
\affiliation{Center for Theoretical Physics, Polish Academy of Sciences,\\
Aleja Lotnik\'ow 32/46, 02-668 Warszawa, Poland}
\affiliation{Guangdong Technion - Israel Institute of Technology, 241 Daxue Road,
Jinping District, Shantou, Guangdong Province, China}

\begin{abstract}

We introduce a measure for evaluating the efficiency of finite universal quantum gate sets $\mathcal{S}$, called the Quantum Circuit Overhead (QCO), and the related notion of $T$-Quantum Circuit Overhead ($T$-QCO). QCO compares the circuit length required by $\mathcal S$ with the best possible length among gate sets of the same size. The $T$-QCO adapts this idea to cost models in which only selected costly gates are counted, while cheap operations are absorbed into an effective gate set. We demonstrate the usefulness of the ($T$-)QCO by extensive numerical calculations of its upper bounds, providing insight into the efficiency of various choices of single-qubit $\mathcal{S}$, including Haar-random gate sets and the gate sets derived from finite subgroups, such as Clifford and Hurwitz groups. In particular, our results suggest that, in terms of the upper bounds on the $T$-QCO, the famous T gate is a highly non-optimal choice for the completion of the Clifford gate set, even among the gates of order 8. We identify the optimal choices of such completions for both finite subgroups.

\end{abstract}

\maketitle

\section{Introduction}
A quantum circuit \cite{Nielsen_Chuang_2010, Kitaev_Yu_Shen_2002} is a universal model for quantum computation in which quantum information is processed via the application of a series of unitary operations called quantum logic gates.
Similarly to a classical computer, whose computation can be described using the classical circuit model, every global quantum operation on a qubit register can be realized using a universal finite set of elementary operations. A set of such quantum logic gates is referred to as the universal gate set or, in the context of quantum hardware, the native gate set.

Contrary to the classical case, finite-length quantum circuits built from a finite discrete set $\mathcal{S}$ of quantum gates can implement arbitrary multi-qubit (global) unitary operations only approximately, up to some error $\epsilon$ (in a suitable metric). The number of elementary gates needed to implement a target unitary operation $U$ with precision $\epsilon$ using gates from $\mathcal{S}$ is a measure of the complexity of $U$ with respect to $\mathcal{S}$ \cite{Nielsen_Chuang_2010, Kitaev_Yu_Shen_2002, aaronson2016complexityquantumstatestransformations}. For a universal gate set $\mathcal{S}$ and any finite $\epsilon$, the complexity of any $U$ is finite and thus can be upper bounded by the shortest circuit length, $\ell (\mathcal{S}, \epsilon)$, so that any $U$ can be $\epsilon$-approximated by a quantum circuit built out of $\mathcal{S}$ of length at most $\ell (\mathcal{S}, \epsilon)$. This number can be understood as an absolute measure of the efficiency of $\mathcal{S}$ at the scale of $\epsilon$-approximations. Since the implementation of quantum gates is always flawed, for reasonably small nonzero $\epsilon$, this number fully characterizes the efficiency of $\mathcal{S}$.

Quantum compilation \cite{ge2024quantumcircuitsynthesiscompilation,H_ner_2018, Nielsen_Chuang_2010} is a process whose main objective is to approximate the target quantum circuit from the high-level hardware-agnostic representation used by quantum programmers to the form expressible by the native gate set executable on a specific quantum computer. Another task handled by the compiler is circuit optimization, which, loosely speaking, involves reducing the resources of quantum circuits, such as the depth of the circuit or the number of specific gates used.  In the case of the current noisy intermediate-scale quantum (NISQ) machines, which do not enjoy quantum error correction, the reduction of the circuit depth and the number of costly gates (such as the noisy entangling gates) is of particular practical importance \cite{Gheorghiu_2023, Preskill2018quantumcomputingin, Noh_2020}. On the other hand, in the fault-tolerant regime, due to the Eastin-Knill theorem \cite{PhysRevLett.102.110502, Woods_2020, Faist_2020}, the number of resource-costly non-transversal gates often determines the bottleneck \cite{gottesman2005quantumerrorcorrectionfaulttolerance, doi:10.1073/pnas.2026250118,Eastin_2009}.
For example, in the case of Clifford+T gate sets realized using many topological codes, such as 2D surface or color codes, the focus is usually on the reduction of the T-count, i.e., the number of non-transversal T gates (also known as the $P(\pi/4)$ or $\pi/8$ gates \footnote{To avoid confusion with the $T$ symbol occurring in $T$-QCO, we refer to the T gate as $P(\pi/4)$ gate.}), which leads to an improvement in error rates, runtime, and the number of qubits needed to perform the computations \cite{PRXQuantum.2.020341, Gheorghiu_2022, Ruiz2025, 10.5555/2685179.2685180,vandaele2024lowertcountfasteralgorithms,zhou2024algorithmicfaulttolerancefast, heyfron2018efficientquantumcompilerreduces,Fowler_2009}. However, the compilation process is fundamentally limited by the efficiency of the used gate set $\mathcal{S}$.

Aside from the applications in the description of information processing occurring in quantum computers, quantum circuits can be used to describe the discrete unitary dynamics of general discrete quantum systems \cite{Tokusumi_2018, Fisher_2023, Claeys_2022}. Such an approach has been recently proposed to gain insight into the physics of black hole interiors, and interesting results regarding the saturation and recurrence of the complexity of such systems have been obtained \cite{Hayden_2007, PhysRevX.14.041068}. Such behavior also depends on the efficiency of gate sets $\mathcal{S}$ used to model the system.

Although it is conjectured that the generic universal gate sets $\mathcal{S}$ have a so-called spectral gap, which implies the optimal asymptotic efficiency $\ell(\mathcal{S}, \epsilon) = \Theta(\log(1/\epsilon))$, the quantitative methods to bound and compare the efficiency of various gate sets $\mathcal{S}$ are not well-developed.

In this work, we introduce and study the relative measure of the efficiency of universal gate sets $\mathcal{S}$ that we call the quantum circuit overhead (QCO) and the related notion of $T$-Quantum Circuit Overhead ($T$-QCO).
The notion of overhead is based on the comparison of
the efficiency $\ell(\mathcal{S}, \epsilon)$ among the gate sets $\mathcal{S}$ having the same number of elements, where the optimal efficiency is denoted $\ell_{\mathrm{opt}}(|\mathcal{S}|, \epsilon)$.
Crucially, both overheads can be upper-bounded by essentially calculable quantities, namely $Q$ and $Q_T$, respectively, which can be obtained from numerical simulations.

To demonstrate the feasibility of our method and its applications, we provide extensive numerical examples in which we calculate $Q/Q_T$, focusing on the comparison between the two scenarios for single-qubit gate sets:
\begin{enumerate}
    \item A Haar-random set $\mathcal{S}$ with a fixed number of elements (of infinite or fixed finite order $r$),
    \item A set $\mathcal{S}$ composed of a finite group (such as Clifford or Hurwitz group) completed with a single Haar-random gate (of infinite or fixed finite order $r$), making the set universal.
\end{enumerate}
In the second scenario, we compare such random ensembles with some ``special'' choices, e.g. the $P(\pi/4)$ gate in the case of the Clifford group, gaining insight into their efficiency. The inclusion of the finite order cases is motivated by the fault-tolerance considerations and the analysis of the so-called Super-Golden Gates \cite{sarnak2015}. Surprisingly, our results suggest that the $P(\pi/4)$ gate is a highly non-optimal choice among all gates of order $r=8$ in terms of $Q_T$. We also identify the best possible gates of orders $r=8$ and $r=2$ in the Clifford and Hurwitz group cases, respectively.

Although our numerical experiments focus on the single-qubit setting, our framework can be applied in any dimension, in particular to multi-qubit gate sets. Moreover, it can in principle be applied to settings in which the universal gate set is not discrete, e.g., when it consists of parametrized gates. We refrain from performing such experiments here because of their computational cost.

To upper bound the overhead, we need to be able to upper bound $\ell(\mathcal{S}, \epsilon)$ and lower bound $\ell_{\mathrm{opt}}(|\mathcal{S}|, \epsilon)$.

\section{Solovay--Kitaev-like theorems}
Lossless unitary quantum operations on an $n$-qubit register are described via the unitary channels $\mathbf{U}(\rho) = U \rho U^{\dagger}$, which form a group $\mathbf{U}(d)$, where $d=2^n$. This group is naturally identified with the projective unitary group $\mathrm{PU}(d)$. We use the following metric on $\mathbf{U}(d)$:
\begin{align}
\label{eq:d_m}
    d(\mathbf{U}, \mathbf{V}) &\coloneqq \min_{\varphi}||U-e^{i \varphi}V||_{\infty} \mathrm{,}
\end{align}
where $||\cdot||_{\infty}$ denotes the operator norm and $U$ and $V$ are unitary representatives of the channels $\mathbf{U}$ and $\mathbf{V}$, respectively (see Appendix \ref{app:unitary} for more details).

The famous Solovay--Kitaev (SK) theorem states that if $\mathcal{S} \subset \mathbf{U}(d)$ is a finite universal symmetric (i.e. inverse-closed) set of quantum gates, then $\ell(\mathcal{S}, \epsilon)=\mathcal{O}(\log^{c}(1/\epsilon))$, where the constant $c$ depends on the proof and is typically $c\approx 3.97$ or $c=3 +\alpha$, for any $\alpha>0$ \cite{dawson2005solovaykitaevalgorithm, Kitaev_Yu_Shen_2002, Nielsen_Chuang_2010}. The proofs are constructive, so that an (efficient) algorithm exists that can find the desired decompositions. As a result, the SK algorithm has served as the foundation of modern quantum compilation. Since its introduction, many similar (constructive and non-constructive) poly-logarithmic upper bounds $\ell(\mathcal{S}, \epsilon)=\mathcal{O}(\mathrm{poly}(\log(1/\epsilon)))$ have been provided \cite{kuperberg2023breakingcubicbarriersolovaykitaev,bouland2021efficientuniversalquantumcompilation, Harrow_2002,S_owik_2023, Dolgopyat2002, varju13, 9614165,słowik2025fundamentalsolutionsheatequation}. Such theorems often work for groups other than $\mathbf{U}(d)$, e.g., semi-simple compact Lie groups, and use different assumptions on the gates in $\mathcal{S}$; we refer to them as Solovay--Kitaev-like (SKL) theorems.

For example, in terms of constructive/algorithmic SKL theorems, the cubic $\ell(\mathcal{S}, \epsilon)$ scaling in the SK algorithm was recently improved in \cite{kuperberg2023breakingcubicbarriersolovaykitaev} to $\log_{\phi}(2) \approx 1.44$, where $\phi$ is the golden ratio. The construction assumes that $\mathcal{S}$ is finite and inverse-closed. On the other hand, in \cite{bouland2021efficientuniversalquantumcompilation}, the authors provided the generalization of the SK algorithm working for any finite universal (i.e., not necessarily inverse-closed) sets $\mathcal{S}$, with $\ell(\mathcal{S}, \epsilon)=\mathcal{O}(\log^{\gamma_d}(1/\epsilon))$ and $\gamma_d=\Theta(\log(d))$.

However, it is known that for finite $\mathcal{S}$, all poly-logarithmic bounds with exponent 1 are asymptotically tight. The Haar volume\footnote{Due to translational invariance of Haar measure and the metric, the volume of a ball does not depend on its origin.} of an $\epsilon$-ball $ B_{\epsilon}\subset \mathbf{U}(d)$ can be bounded as
\begin{equation}
\label{eq:volb}
 (a_v\epsilon)^{d^2-1}\leq \mathrm{Vol}(B_\epsilon)\leq (A_v\epsilon)^{d^2-1} \mathrm{,}
\end{equation}
with known constants $a_v=\frac{1}{9 \pi}$ and $A_v=87$. These constants were provided in \cite{9614165} using methods from \cite{szarek1998}. Then, using the simple volume counting argument \cite{Harrow_2002, Nielsen_Chuang_2010}, one may express the lower bound on $\ell(\mathcal{S}, \epsilon)$ as
\begin{equation}
\label{eq:lv}
   \ell_{\mathrm{vol}}(|\mathcal{S}|, \epsilon) \approx  \frac{d^2-1}{\log(|\mathcal{S}|)}\log\left(\frac{1}{A_v \epsilon}\right) \mathrm{,}
\end{equation}
where $A_v=87$. Thus, $\ell(\mathcal{S}, \epsilon) = \Omega(\log(1/\epsilon))$.

This lower bound depends only on the number of elements of $\mathcal{S}$. Hence, it can be used to lower bound $\ell_{\mathrm{opt}}(|\mathcal{S}|, \epsilon)$, which yields

\begin{equation}
\label{eq:lvb}
    \ell(\mathcal{S}, \epsilon) \geq \ell_{\mathrm{opt}}(|\mathcal{S}|,\epsilon) \geq  \ell_{\mathrm{vol}}(|\mathcal{S}|, \epsilon)
\end{equation}

Such an optimal scaling $\Theta(\log(1/\epsilon))$ can be obtained when $\mathcal{S}$ has a so-called spectral gap. It is useful to reformulate this property to the language of unitary $\delta$-approximate $t$-designs.

A unitary $\delta$-approximate $t$-design is a probability measure $\nu$ on $\mathbf{U}(d)$ that mimics the averaging properties of Haar measure $\mu$ when applied to balanced polynomials with degree bounded by $t$, up to a discrepancy $\delta(\nu,t):=\left\|T_{\nu,t}-T_{\mu,t}\right\|_{\infty}$,
where
\begin{equation}
\label{eq:t_moment}
     T_{\mu,t}\coloneqq\int_{\mathbf{U}(d)} d\mu(U) U^{ t,t},\,\,\,\,\,T_{\nu,t}\coloneqq \int_{\mathbf{U}(d)} d\nu(U) U^{ t,t} \mathrm{,}
\end{equation}
are so-called $t$-moment (averaging) operators, $U^{t,t}:=U^{\otimes t}\otimes \bar{U}^{\otimes t}$, and we require $\delta(\nu,t) < 1$ (see Appendix \ref{app:t_designs} for more information). For any gate set $\mathcal{S}$, by $\nu_{\mathcal{S}}$ we denote the uniform probability measure supported on its elements.

Note that for symmetric $\mathcal{S}$,
\begin{equation}
||T^{\ell}_{\nu_{\mathcal{S}},t}-T_{\mu,t}||_{\infty} = \delta^{\ell}(\nu_{\mathcal{S}},t)
\end{equation}
quantifies the difference between averaging over circuits of length $\ell$ and averaging with respect to Haar measure \footnote{For non-symmetric $\mathcal{S}$ we have an inequality.}. Therefore, the smaller $\delta(\nu_{\mathcal{S}},t)$ is, the shorter the circuits needed to mimic the Haar averaging.

The spectral gap of $\mathcal{S}$ is then $1- \delta(\nu_\mathcal{S})$, where $\delta(\nu_\mathcal{S})$ is the supremum of $\delta(\nu_\mathcal{S},t)$ over all scales $t$, so that the spectral gap property reads $\delta(\nu_{\mathcal{S}}) < 1$. The quantitative version of the statement about the efficiency of gate sets $\mathcal{S}$ with a spectral gap is a non-constructive SKL theorem  \cite{Harrow_2002,S_owik_2023} %Essentially the same theorem was proved differently in \cite{S_owik_2023} \footnote{This theorem follows from the proof of Theorem 3   if the bound $(1-x)^r \leq e^{-rx}$ is not used.}
and it states that if $\delta(\nu_\mathcal{S}) <1$, then for any precision $\epsilon$ every operation $U$ from $\mathbf{U}(d)$ can be approximated by a sequence of gates from $\mathcal{S}$ of the length
\begin{equation}
  \frac{d^2-1}{\log\left(1/\delta(\nu_\mathcal{S})\right)}\log\left(\frac{2}{A_v \epsilon}\right) \mathrm{.}
\end{equation}
Notice that although the scaling is optimal, the prefactor may be arbitrarily large. Moreover, in our examples, the prefactor is bounded from below via $\delta(\nu_\mathcal{S})\geq \delta_{\mathrm{opt}}(\mathcal{S})$, where
\begin{equation}
\label{eq:efficient_gates}
    \delta_{\mathrm{opt}}(\mathcal{S}) \coloneqq \frac{2\sqrt{|\mathcal{S}| - 1}}{|\mathcal{S}|} \mathrm{.}
\end{equation} \cite{kesten59}
(see Appendix \ref{app:kesten} for more detailed explanation). We say a finite gate set $\mathcal{S}$ is efficient if $\delta(\nu_\mathcal{S})= \delta_{\mathrm{opt}}(\mathcal{S})$ and refer to $\delta_{\mathrm{opt}}(\mathcal{S})$ as the optimal value. Note that the optimal value depends only on the number of gates $|\mathcal{S}|$.

The study of $\delta(\nu_\mathcal{S})$ for generic $\mathcal{S}$ is a hard problem as $\delta(\nu_\mathcal{S})$ cannot be directly calculated. However, some properties of $\delta(\nu_\mathcal{S})$ are known. For example, it is known that $\delta(\nu_\mathcal{S}) < 1$ for the finite universal sets $\mathcal{S}$ consisting of algebraic elements \cite{Bourgain2007OnTS, bourgain2011spectralgaptheoremsud}. This result was later generalized to any compact, simple Lie group  \cite{benoist2014spectralgaptheoremsimple}. Moreover, it has been conjectured (and is now commonly believed) that $\delta(\nu_\mathcal{S}) < 1$ for any finite universal $\mathcal{S}$ and there are known examples of efficient finite single-qubit gate sets $\mathcal{S}$ with $|\mathcal{S}|= p + 1$ for $p \equiv 1 \, \mathrm{mod} \, 4$ \cite{lps86, lps87}. Finally, some commonly used one-qubit gate sets are known to be efficient \cite{bocharov2013, selinger2015, sarnak2015, kliuch2016}. To the best of our knowledge, the construction of efficient many-qubit gates remains an open problem.

Fortunately, one can still obtain useful non-constructive SKL theorems using the knowledge of $\delta(\nu_\mathcal{S},t)$. Such a finite-scale approach was studied in \cite{Dolgopyat2002, varju13, 9614165, słowik2025fundamentalsolutionsheatequation} and is sufficient in practice, as it corresponds to studying efficiency at a certain finite precision $\epsilon$. The approach from \cite{9614165, słowik2025fundamentalsolutionsheatequation} utilizes the relation between $\epsilon$-nets and $\delta$-approximate $t$-designs.

A subset of channels $\mathcal{E}$ from $\mathbf{U}(d)$ is an $\epsilon$-net if for every channel $\mathbf{U}$ from $\mathbf{U}(d)$, there exists a channel $\mathbf{V}$ from $\mathcal{E}$ such that $d(\mathbf{U},\mathbf{V}) \leq \epsilon$. In other words, $\mathcal{E}$ contains representatives of all channels up to error $\epsilon$. It is natural to expect a relation between $\epsilon$-nets formed by circuits over $\mathcal{S}$ and $\delta$-approximate $t$-designs supported on those circuits. However, the quantitative relations between them were not known until recently. Such bounds for the group $\mathbf{U}(d)$ were first rigorously established in \cite{9614165}, where the authors show \footnote{The result is more general as it does not assume that the measure is uniform.} that a set is an $\epsilon$-net if it is a support of a $\delta$-approximate $t$-design with the parameters obeying the following scalings

\begin{equation}
\label{eq:t_eps}
    t(\epsilon)\gtrsim \frac{d^{5/2}}{\epsilon}, \quad \delta(\epsilon) \lesssim \left(\frac{\epsilon^{3/2}}{d}\right)^{d^2}
\end{equation}
 (see \cite{9614165} for precise formulas). A more recent study improves the second scaling to $\delta(\epsilon) \lesssim (\epsilon/d^{1/2})^{d^2}$
 \cite{słowik2025fundamentalsolutionsheatequation}.

From the viewpoint of nonabelian Fourier analysis on groups, this reciprocal relation between $t$ and $\epsilon$ can be intuitively understood as the relation between distances on the group and the corresponding ``frequency'' space, so that smaller $\epsilon$ corresponds to faster-varying functions.
 A quantitative version of this SKL theorem was proved in \cite{9614165} and states \footnote{Proposition 2 in \cite{9614165} has $1- \delta(\nu_{\mathcal{S}},t)$ instead of $\log\left(1/\delta(\nu_\mathcal{S},t)\right)$ because of an unnecessary bound.} that for a fixed precision $\epsilon$, every operation $U$ from $\mathbf{U}(d)$ can be $\epsilon$-approximated by sequences of gates from $\mathcal{S}$ of the length $\ell_{\delta}(\mathcal{S}, \epsilon)$
\begin{equation}
\label{eq:ub}
     \ell(\mathcal{S}, \epsilon)  \leq \ell_{\delta}(\mathcal{S}, \epsilon) \sim \frac{d^2-1}{\log\left(1/\delta(\nu_\mathcal{S},t(\epsilon))\right)}\log\left(\frac{1}{\epsilon}\right) \mathrm{,}
\end{equation}
where $t(\epsilon)$ is the bound of type (\ref{eq:t_eps}) stemming from the $\epsilon$-net $t$-design correspondence. Thus, we can say that $\delta(\nu_\mathcal{S}, t(\epsilon))$ upper bounds the efficiency of $\mathcal{S}$ on the level of $\epsilon$-approximations. Moreover, for not too large values of $t$ and $d$, the value of $\delta(\nu_\mathcal{S},t)$ can be calculated using supercomputing clusters. Conveniently, contrary to the Solovay-Kitaev theorem, such SKL theorem can be applied to arbitrary $\mathcal{S}$, in particular to continuous $\mathcal{S}$.

The distribution of $\delta(\nu_\mathcal{S},t)$ for (fully) Haar-random ensembles of finite $\mathcal{S}$ was studied in \cite{Dulian_2023}, with the extensive numerical analysis suggesting fast stabilization of the distribution with growing $t$. Our numerical experiments further validate this observation and extend it to all types of ensembles of gate sets studied in this paper. Hence, although the bounds (\ref{eq:t_eps}) provide some theoretical guarantees on the scales $t$ needed to gain insight into the $\epsilon$-scale efficiency (via (\ref{eq:ub})), our results suggest that in practice, it suffices to compute $\delta(\nu_\mathcal{S},t)$ for $t$ much smaller than the bounds $t(\epsilon)$.

Although Eq.~(\ref{eq:ub}) suggests that $\delta(\nu_\mathcal{S}, t)$ is a good measure of the efficiency of finite $\mathcal{S}$, the value of $\delta(\nu_\mathcal{S}, t)$ is sensitive to the number of gates $|\mathcal{S}|$. In particular, as the number of gates $|\mathcal{S}|$ goes to infinity, the optimal value (\ref{eq:efficient_gates}), which lower bounds the supremum of $\delta(\nu_\mathcal{S}, t)$ over $t$, goes to $0$.
Since the implementation of gate sets $\mathcal{S}$ with large $|\mathcal{S}|$ is costly in practice, e.g. due to the necessary calibrations of quantum hardware, it makes sense to compare the gate sets $\mathcal{S}$ of fixed $|\mathcal{S}|$. This motivates us to introduce the notion of the overhead of quantum circuits.

\section{Quantum Circuit Overhead}
\label{sec:qco}
We define the Quantum Circuit Overhead (QCO) of a finite universal gate set $\mathcal{S}$ for $\epsilon$-approximations as the ratio between the smallest length of circuits over $\mathcal{S}$ which form an $\epsilon$-net, $\ell(\mathcal{S},\epsilon)$, and the optimal length $\ell_{\mathrm{opt}}(|\mathcal{S}|,\epsilon )$ achievable using gate sets with the same number of gates $|\mathcal{S}|$. This quantity is very hard to calculate in general; however, we can bound it from above by bounding $\ell(\mathcal{S},\epsilon)$ from above and $\ell_{\mathrm{opt}}(|\mathcal{S}|,\epsilon )$ from below using (\ref{eq:lv}), (\ref{eq:lvb}) and (\ref{eq:ub}) as follows:

\begin{equation}
    \frac{\ell(\mathcal{S}, \epsilon)}{\ell_{\mathrm{opt}}(|\mathcal{S}|, \epsilon)} \leq \frac{\ell_{\delta}(\mathcal{S}, \epsilon)}{\ell_{\mathrm{vol}}(|\mathcal{S}|, \epsilon)} \lesssim Q(\mathcal{S},\epsilon) \mathrm{,}
\end{equation}
where we define the computable upper bound on QCO as
\begin{equation}
\label{eq:Q}
Q(\mathcal{S}, \epsilon) \coloneqq \frac{\log(|\mathcal{S}|)}{\log\left(1/\delta(\nu_\mathcal{S},t(\epsilon))\right)} \mathrm{,}
\end{equation}
and $t(\epsilon)$ is the bound stemming from the $\epsilon$-net $t$-design correspondence of type (\ref{eq:t_eps}). Note that $Q(\mathcal{S},\epsilon)$ is a non-increasing function of $\epsilon$. It is interesting to study the asymptotic behavior of (\ref{eq:Q}) in the limit of $\epsilon \to 0$ (i.e. $t \to \infty$), namely we define
\begin{equation}
\label{eq:Q_o}
    \overline{Q}(\mathcal{S}) \coloneqq \limsup_{\epsilon\rightarrow 0}  Q(\mathcal{S}, \epsilon) \mathrm{.}
\end{equation}
For efficient gate sets, we can use (\ref{eq:efficient_gates}) to obtain
\begin{equation}
\overline{Q}_{\mathrm{opt}}(\mathcal{S})\coloneqq\frac{\log(|\mathcal{S}|)}{\log\left(\frac{|\mathcal{S}|}{2\sqrt{|\mathcal{S}|-1}}\right)} \geq 2 \mathrm{,}
\end{equation}
where $\overline{Q}_{\mathrm{opt}}(\mathcal{S})\gtrsim 2$ for large $|\mathcal{S}|$. We refer to $\overline{Q}_{\mathrm{opt}}(\mathcal{S})$ as the optimal value, since it is a lower bound on $\overline{Q}(\mathcal{S})$ attainable on the efficient gate sets $\mathcal{S}$.

Notably, our definition of QCO still makes sense for infinite $\mathcal{S}$, but then it simplifies to the efficiency $\ell(\mathcal{S}, \epsilon)$ due to $\ell_{\mathrm{opt}}(|\mathcal{S}|, \epsilon)=1$ being realized trivially.

The notion of QCO is suitable for scenarios in which one is interested in the pure computational efficiency of the gate sets or, in the context of quantum computers, the total gate count of the circuits (see Example \ref{exa:1}). Practical architectures in which such a scenario may be relevant include the homogeneous-cost models based on anyons (see Table \ref{tab:use_cases}).

\section{\texorpdfstring{$T$}{T}-Quantum Circuit Overhead}
In many architectures, physical gate costs have a clear hierarchy: some gates are costly, while other operations are much cheaper. The $T$-QCO is a first-order proxy for this situation. It counts only the costly gates and neglects the cost of the cheap operations. Thus the cheap operations are ``free'' only in the sense of this counting approximation, not in the literal physical sense.

Let $G_C$ denote a native set of cheap gates and define
\begin{equation}
C=\overline{\langle G_C\rangle}
\end{equation}
as the closed group generated by them. If the generated group is finite, taking the closure does not change it. We consider gate sets of the form
\begin{equation}
\label{eq:S_ansatz}
\mathcal{S} = C \cup \left\{ T_1, \ldots, T_k \right\} \mathrm{,}
\end{equation}
where the gates $T_i\notin C$ are the costly gates whose uses are counted. The unweighted $T$-QCO assumes that the costs of the $T_i$ are comparable, while implementing elements of $C$ is negligible on that scale. The role of $G_C$ and the corresponding diameter assumption are discussed in Appendix~\ref{app:T_QCO}.

For finite $C$, one step of the derived random walk represents one use of a costly gate $T_i$, dressed by cheap gates from $C$. This gives the derived gate set
\begin{equation}
\label{eq:derived}
 \mathcal{S}_{T} \coloneqq \left\{ cT_ic^{\dagger}: c \in C,\; i\in[k] \right\} \mathrm{.}
\end{equation}
The word ``derived'' is meant literally: cheap gates have been absorbed into the gates that appear in the random walk. As in ordinary QCO, $\mathcal S_T$ is a set of distinct gates, not a list with repetitions. Removing repetitions does not change the true $T$-complexity, because it does not change which operations can be performed using one costly gate, up to initial or final multiplication by cheap gates; see Appendix~\ref{app:T_QCO}.

For finite $\mathcal{S}_T$, we use the uniform measure on its distinct elements,
\begin{equation}
\label{eq:derived_measure}
 \nu_{\mathcal{S}_T} \coloneqq \frac{1}{|\mathcal{S}_T|}\sum_{g\in \mathcal{S}_T}\delta_g\mathrm{.}
\end{equation}
The corresponding $t$-moment operator controls the convergence of the derived random walk per use of a costly gate. By the same volumetric reasoning as for QCO, the finite-set upper-bound proxy for the $T$-QCO is
\begin{equation}
\label{eq:Q_T}
Q_T(\mathcal{S}, \epsilon) \coloneqq
Q(\mathcal{S}_T,\epsilon)=
\frac{\log|\mathcal{S}_T|}{\log\left(1/\delta(\nu_{\mathcal{S}_T},t(\epsilon))\right)} \mathrm{.}
\end{equation}
The numerator uses $|\mathcal S_T|$, rather than $k|C|$, because the volumetric lower bound depends on the number of distinct gates in the derived random walk. In general,
\begin{equation}
\label{eq:derived_cardinality}
|\mathcal{S}_T| \leq k|C|\mathrm{.}
\end{equation}
Strict inequality means that the adjoint action of $C$ has symmetries: some elements of $C$ may stabilize a gate $T_i$, or two gates $T_i$ and $T_j$ may lie in the same adjoint orbit. Such repetitions are removed in $\mathcal S_T$; otherwise the volumetric comparison would artificially penalize a reducible description of the same effective operations. Details are given in Appendix~\ref{app:T_QCO}.

For continuous cheap gate groups, the moment-operator framework above already applies with integrals instead of finite sums. If the derived set is continuous, then, as for ordinary QCO with infinite gate sets, the exact optimal denominator is $1$ and the relevant object is the complexity, or its spectral upper bound, for the chosen measure. The finite formula in Eq.~\eqref{eq:Q_T} is therefore used only after choosing a finite subgroup, a finite derived set, or a finite compilation model. Finally, the $T$-QCO upper bound is well-defined for reasonably small $\epsilon$, so that the denominator is nonzero.

In contrast to QCO, the notion of $T$-QCO is most suitable when the gate set has a clear cheap/costly split. For example, in NISQ architectures, $C$ may describe cheap local controls and the $T_i$ may be native entangling gates (see Example \ref{exa:3}). In fault-tolerant or QEC architectures, $C$ is often a Clifford group, while the $T_i$ are non-Clifford resources such as magic-state-injected phase gates or CCZ-type gates (see Table \ref{tab:use_cases} and Example \ref{exa:4}).

\section{Applications and cost models}
\label{sec:applications}

We recall that QCO applies to universal gate sets with a homogeneous cost model. The $T$-QCO applies when the gate set has the decomposed form in Eq.~\eqref{eq:S_ansatz}: $C$ is generated by cheap native operations and only the gates $T_i$ are included in the cost count. Below, we give three representative scenarios. Table~\ref{tab:use_cases} summarizes architectures where either QCO or $T$-QCO can be used as a first-order proxy for efficiency.

\begin{exa}[single-qubit gate count]
\label{exa:1}
Consider a single-qubit NISQ architecture with a gate set $\mathcal{S}$, consisting of gates with similar fidelities. Then the QCO of $\mathcal{S}$, which boils down to the analysis of the gate count/circuit depth, is a sensible measure of the efficiency of $\mathcal{S}$.
\end{exa}

\begin{exa}[CNOT-count flavor]
\label{exa:3}
Consider a NISQ $n$-qubit architecture with parametrized 2-qubit entangling gates $\mathrm{Ent}_{i, i+1}(\overline{\phi})$ with similar fidelities, acting on qubits $i$ and $i+1$ for $1 \leq i \leq n-1$. We pick $\mathcal{S}$ as in \eqref{eq:S_ansatz}, where $C=\mathbf{U}(2)^{\otimes n}$ is the cheap local-control group and $T_i = \mathrm{Ent}_{i, i+1}(\overline{\phi})$, for $1 \leq i \leq n-1$ \footnote{Here $\mathbf{U}(2)$ is used to integrate out single-qubit operations and focus on the impact of the entangling gates. A finite single-qubit gate set can also be used if its compilation cost is negligible compared with the entangling gates.}. Then the $T$-QCO of $\mathcal{S}$ is a sensible measure of efficiency with respect to the choice of $\overline{\phi}$.
\end{exa}

\begin{exa}[T-count flavor]
\label{exa:4}
Consider a fault-tolerant architecture with $n$ logical qubits, such that the Clifford gates are low-cost compared to the parametrized family of non-Clifford phase gates $P(\phi)$, which can be implemented with similar cost. We pick $\mathcal{S}$ as in \eqref{eq:S_ansatz}, where $C=\mathcal{C}_n$ is the $n$-qubit Clifford group and $T_i$, for $1 \leq i \leq n$, is the non-Clifford $P(\phi)$ gate acting on the $i$-th qubit. Then the $T$-QCO of $\mathcal{S}$ is a sensible measure of efficiency with respect to the choice of $\phi$.
\end{exa}

\begin{table*}[t]
\centering
\footnotesize
\setlength{\tabcolsep}{3pt}
\renewcommand{\arraystretch}{1.15}
\caption{\label{tab:use_cases} Examples of architectures to which ($T$-)QCO can be applied as a reasonable proxy for overall efficiency.}
\begin{tabular}{|p{1.6cm}|p{2.4cm}|p{3.0cm}|p{3.2cm}|p{2.8cm}|p{1.2cm}|}
\hline
\textbf{Category} & \textbf{Architecture} & \textbf{Cheap operations ($C$)} & \textbf{Costly operations ($\{T_i\}$)} & \textbf{Cost split} & \textbf{Metric} \\
\hline
NISQ & NISQ devices (general) \cite{Preskill2018quantumcomputingin} & Local single-qubit group on each qubit (Euler/ZXZ gates not counted) & Entangling gates (e.g. CNOT/CZ/MS/Rydberg) & \textbf{High}; 2-qubit gates dominate time/error & \textbf{T-QCO} \\
\hline
\multirow{5}{*}{\shortstack[l]{Fault-tolerant\\(Code-based)}}
& 2D surface code \cite{Horsman_2012, Litinski_2019} & Clifford subgroup via lattice surgery / Pauli-based computation & T via magic state distillation (factory limited) & \textbf{Very high}; T-state throughput bottleneck & \multirow{5}{*}{\textbf{T-QCO}} \\
\cline{2-5}
& 2D color code \cite{PhysRevA.76.012305,bombin2015gaugecolorcodesoptimal} & Transversal Clifford subgroup (baseline codes) & T via magic state distillation or gauge-fixing / code-switching & \textbf{Very high}; T preparation dominates &  \\
\cline{2-5}
& 3D surface code \cite{PhysRevA.100.012312} & Subgroup generated by Cliffords \emph{and} transversal CCZ (treat CCZ as cheap) & T via magic state distillation & \textbf{High}; cheap CCZ leaves T as the main costly gate &  \\
\cline{2-5}
& 3D color code (baseline) \cite{PhysRevA.91.032330,bombin2015gaugecolorcodesoptimal} & Transversal Clifford + CCZ subgroup (cheap); some boundary/gauge variants enable transversal T & T via injection/gauge-fixing (when not made transversal) & \textbf{High}; clear cheap/costly split in baseline setting &  \\
\cline{2-5}
& Triorthogonal codes / CSS-T (factories) \cite{PhysRevA.86.052329} & Clifford subgroup inside distillation circuits & Production/consumption of high-fidelity T/CCZ resource states & \textbf{Very high} within factory; resource states dominate &  \\
\hline
\multirow{2}{*}{\shortstack[l]{Fault-tolerant\\(Anyonic)}}
& Ising / Majorana anyons \cite{RevModPhys.80.1083,PhysRevLett.104.180505} & Clifford subgroup by braiding & T via magic state injection (or equivalent) & \textbf{High}; injections dominate & \textbf{T-QCO} \\
\cline{2-6}
& Fibonacci anyons (braiding-universal) \cite{RevModPhys.80.1083} & No robust cheap subgroup; all gates from braids & All gates via braiding (cost by compiled braid length) & \textbf{Homogeneous cost}; no cheap/costly split & \textbf{QCO} \\
\hline
\end{tabular}
\end{table*}

\section{Numerical examples}
\label{sec:ex}
We provide numerical examples focusing on the calculation of the upper bounds on QCO and $T$-QCO, given by $Q$ (\ref{eq:Q}) and $Q_T$ (\ref{eq:Q_T}), respectively  (see Appendix \ref{app:num_expl} for more details about the methods used in numerical experiments). The calculations were performed on a supercomputing cluster.

We consider two types of finite universal one-qubit gate sets:
\begin{enumerate}
    \item Haar-random gate sets with $n$ elements of (finite or infinite) order $r$, denoted $\mathcal{S}_{\mu, n, r}$,
    \item gate sets derived from a finite subgroup $C \subset \mathbf{U}(2)$ generated by a cheap set $G_C$:
    \begin{enumerate}
        \item completed with a fixed costly gate $T$, denoted $C_T$,
        \item completed with a single Haar-random costly gate of (infinite or fixed finite) order $r$, denoted $C_{\mu,r}$,
    \end{enumerate}
    following the setting (\ref{eq:S_ansatz}).
\end{enumerate}

\begin{figure}
  \centering
      \includegraphics[width=0.5\textwidth]{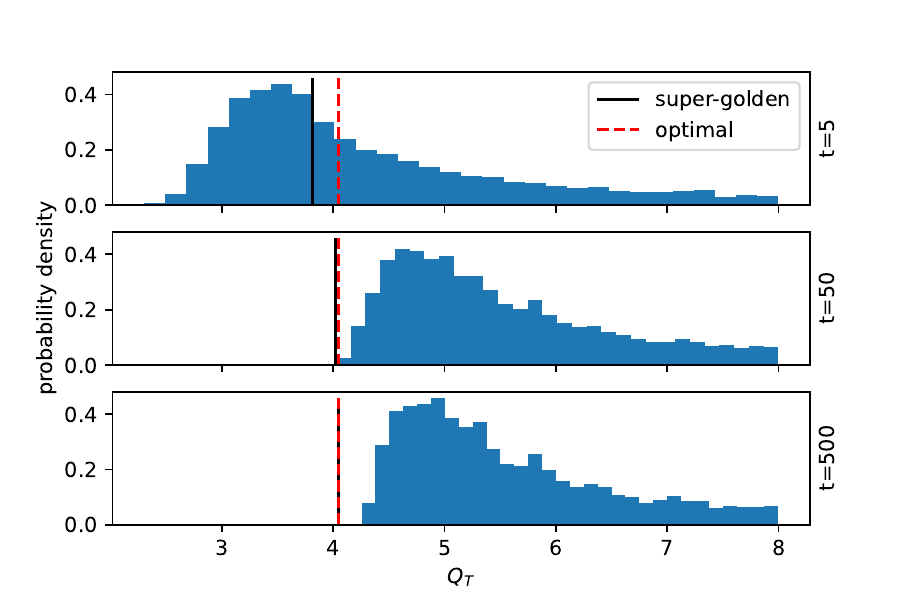}
  \caption{Histograms of the $Q_T$ probability density for an ensemble of type $\mathcal{H}_{\mu, \infty}$ with increasing $t$. The dashed line denotes the corresponding optimal value. The solid line corresponds to a Super-Golden gate set $\mathcal{H}_{T_{12}}$.}
  \label{fig:hurwitz_QT_t5-50-500_sg}
\end{figure}

\begin{figure}
  \centering
      \includegraphics[width=0.5\textwidth]{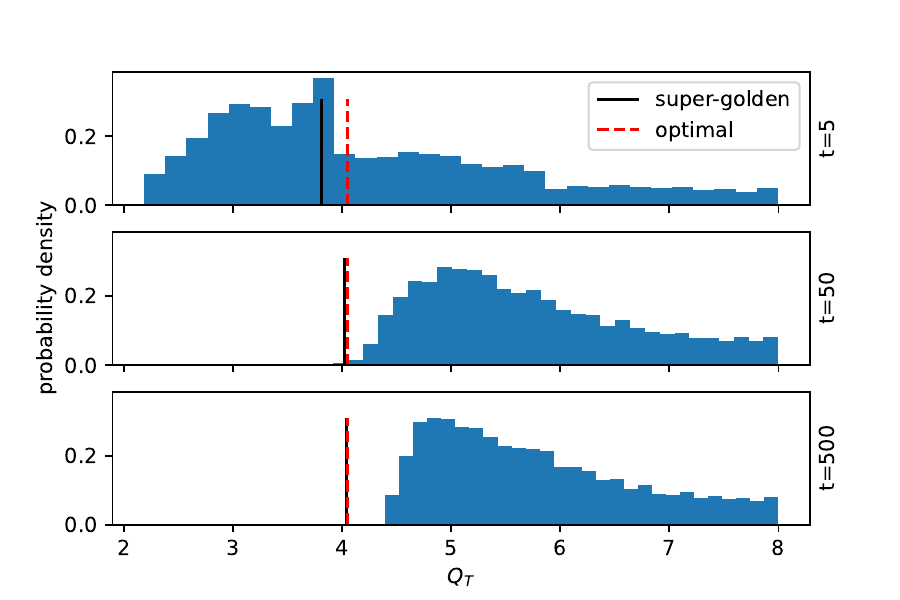}
  \caption{Histograms of the $Q_T$ probability density for an ensemble of type $\mathcal{H}_{\mu,2}$ with increasing $t$. The dashed line denotes the corresponding optimal value. The solid line corresponds to a Super-Golden gate set $\mathcal{H}_{T_{12}}$.}
  \label{fig:hurwitz_QT_t5-50-500_r2_sg}
\end{figure}

We analyze two choices of one-qubit cheap gate group $C$: the Clifford group $\mathcal{C}=\langle G_{\mathcal C}\rangle$ and the Hurwitz group $\mathcal{H}=\langle G_{\mathcal H}\rangle$. In the numerical computation of $Q_T$, the relevant object is the reduced derived set $\mathcal S_T$ associated with the chosen completion.
For each $C$, we construct a random ensemble of $\approx 10^4$ derived universal gate sets of type $C_{\mu, r}$, where $r$ is $\infty$ or equal to either 8 or 2 for $\mathcal{C}$ and $\mathcal{H}$, respectively. In this way, we obtain histograms representing the probability density of $Q_T$ for a fixed $t$. We increase the value of $t$ until the histograms stabilize and mark the corresponding optimal values of $Q_T$ (see Fig.~\ref{fig:cliff_QT_t5-50-500_sg} and Fig.~\ref{fig:cliff_QT_t5-50-500_r8_sg} for $\mathcal{C}_{\mu,r}$ ensembles and Fig.~\ref{fig:hurwitz_QT_t5-50-500_sg} and Fig.~\ref{fig:hurwitz_QT_t5-50-500_r2_sg} for $\mathcal{H}_{\mu, r}$ ensembles). The optimal value does not depend on the scale $t$ and lower-bounds the histograms in the limit $t \to \infty$.

\begin{figure}
  \centering
      \includegraphics[width=0.5\textwidth]{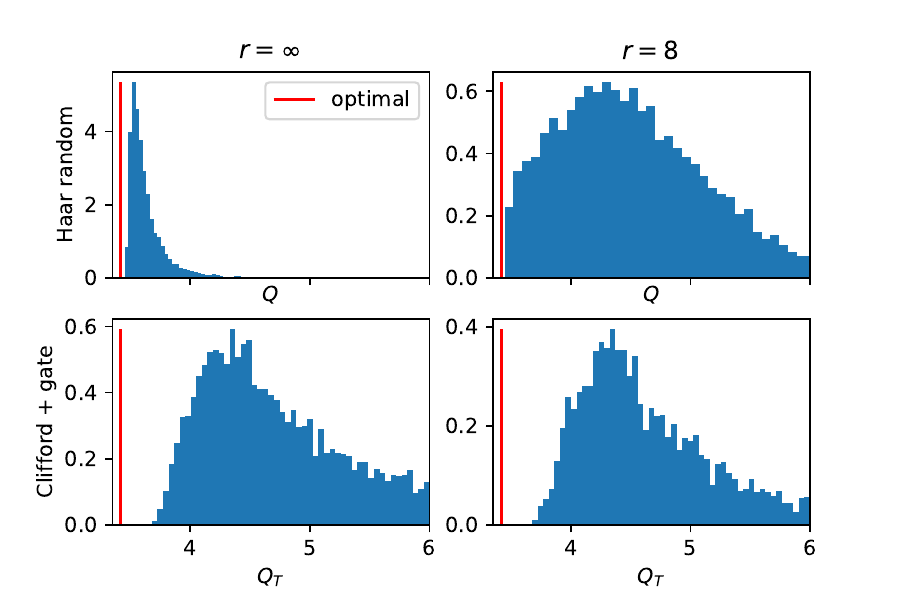}
  \caption{Histograms of the $Q_T$ probability density for ensembles of type $\mathcal{C}_{\mu, r}$ (bottom) versus the histogram of $Q$ for the corresponding ensembles of type $\mathcal{S}_{\mu, 24, r}$ (top) for $t=500$. The solid line denotes the corresponding optimal value. Note that the $y$-axis scales differ.}
  \label{fig:cliff_Q_vs_QT_t500_all4}
\end{figure}

\begin{figure}[t]
  \centering
      \includegraphics[width=0.5\textwidth]{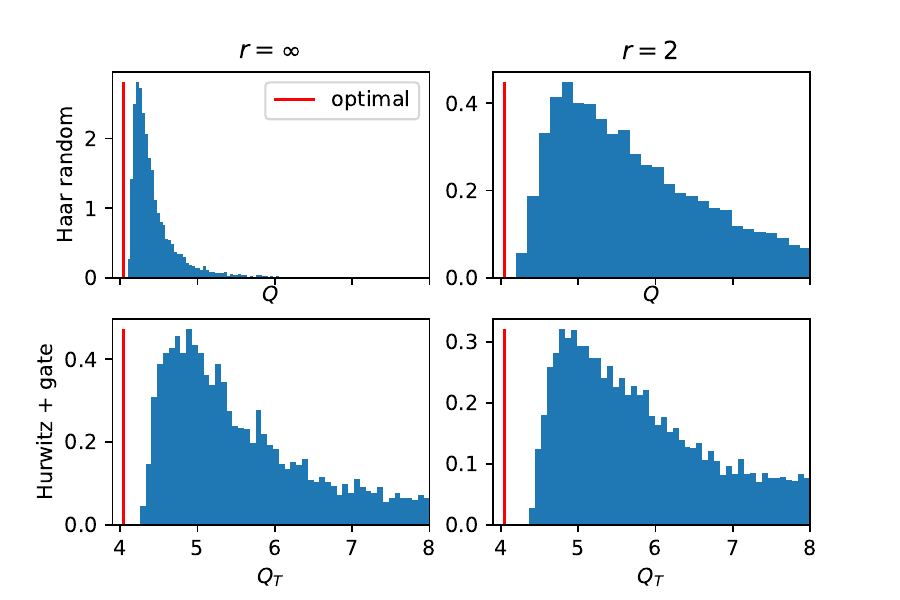}
  \caption{Histograms of the $Q_T$ probability density for ensembles of type $\mathcal{H}_{\mu, r}$ (bottom) versus the histogram of $Q$ for the corresponding ensembles of type $\mathcal{S}_{\mu, 12, r}$ (top) for $t=500$. The solid line denotes the corresponding optimal value. Note that the $y$-axis scales differ.}
  \label{fig:hurwitz_Q_vs_QT_t500_all4}
\end{figure}

Moreover, we compare such histograms with analogous histograms of $Q$ for the same-size ensembles of type $\mathcal{S}_{\mu, n, r}$ containing the corresponding number of gates $n=|C|$ (see Fig.~\ref{fig:cliff_Q_vs_QT_t500_all4} for Clifford group and Fig.~\ref{fig:hurwitz_Q_vs_QT_t500_all4} for Hurwitz group) and with the values of $Q_T$ for gate sets of type $C_T$ with ``special'' choices of $T$.

The comparison with purely random ensembles $\mathcal{S}_{\mu, n, r}$ is relevant from the theoretical point of view, as such gate sets are generic and the distribution of $\delta(\nu_{\mathcal{S}},t)$ can be studied using Random Matrix models \cite{Dulian_2024}.

Finally, we identify choices of $T$ giving the best values of $Q_T$, among all gates of order $r=8$ (for the Clifford group) and $r=2$ (for the Hurwitz group). We achieve this using a Monte Carlo search over the relevant random completions $C_{\mu,r}$.

Additionally, we check the tightness of the bound (\ref{eq:efficient_gates}) in the case of ensembles of type $C_{\mu,r}$ with finite $r$ by calculating the distributions of singular values of the corresponding $t$-moment operator (see Appendix \ref{app:kesten} and Fig.~\ref{fig:cliff_t500_r8_spec} and Fig.~\ref{fig:hurwitz_t500_r2_spec} for more details).

\subsection{Clifford group}
The one-qubit Clifford subgroup $\mathcal{C} \subset \mathbf{U}(2)$ has 24 elements. In the notation of the previous sections, we take the cheap generating set
\begin{equation}
G_{\mathcal C}=\left\{
    \begin{pmatrix}
        1 & 0 \\
        0 &i
    \end{pmatrix},
    \begin{pmatrix}
        1 & 1 \\
        -1 & 1
    \end{pmatrix}
    \right\},
\end{equation}
and $\mathcal{C}=\langle G_{\mathcal C}\rangle$, up to normalization. In the $T$-QCO computation, the Clifford adjoint orbit of the added gate is reduced to distinct elements of $\mathcal S_T$. The special choices of costly $T$ gates include the $P(\pi/4)$ gate (of order $r=8$) and the so-called Super-Golden gate \cite{Parzanchevski_2018} (of order $r=2$), denoted $T_{24}$

\begin{equation}
 P(\pi/4)=
     \begin{pmatrix}
        1 & 0 \\
        0 & 1+i
    \end{pmatrix}, \quad
     T_{24}=
     \begin{pmatrix}
        -1-\sqrt{2} & 2-\sqrt{2}+i \\
        2-\sqrt{2}-i & 1+\sqrt{2}
    \end{pmatrix} \mathrm{,}
\end{equation}
up to normalization.

\begin{figure}
  \centering
      \includegraphics[width=0.5\textwidth]{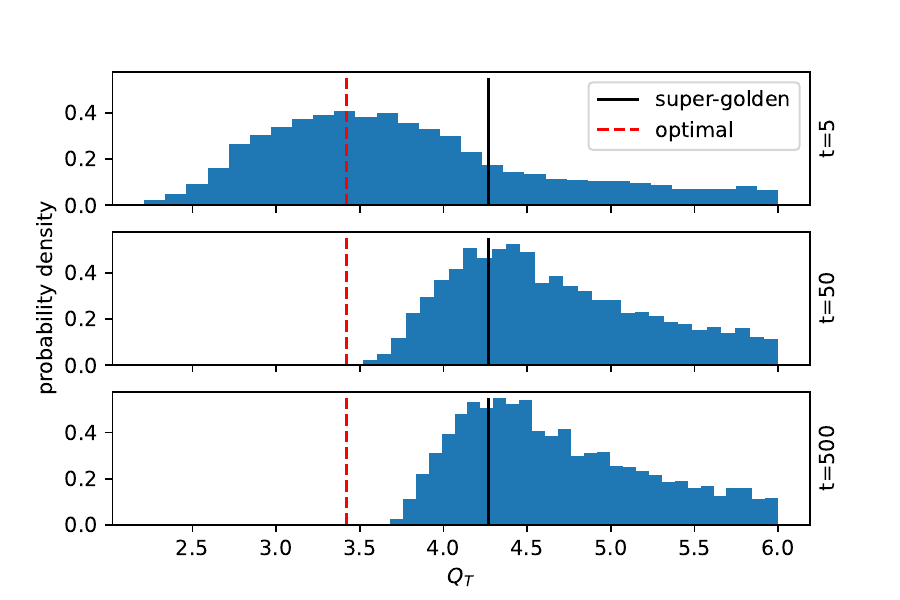}
  \caption{
Histograms of the $Q_T$ probability density for an ensemble of type $\mathcal{C}_{\mu, \infty}$ with increasing $t$. The dashed line denotes the corresponding optimal value. The solid line corresponds to a Super-Golden gate set $\mathcal{C}_{T_{24}}$.}
  \label{fig:cliff_QT_t5-50-500_sg}
\end{figure}

\begin{figure}
  \centering
      \includegraphics[width=0.5\textwidth]{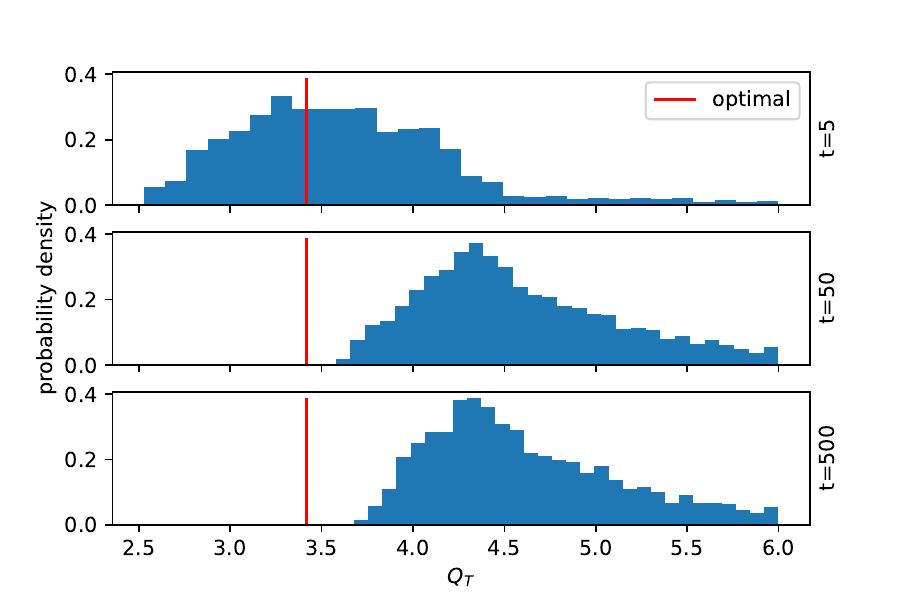}
  \caption{Histograms of the $Q_T$ probability density for an ensemble of type $\mathcal{C}_{\mu,8}$. The solid line denotes the corresponding optimal value.}
  \label{fig:cliff_QT_t5-50-500_r8_sg}
\end{figure}

The value for the gate set $\mathcal{C}_{P(\pi/4)}$ is far outside the range of Fig.~\ref{fig:cliff_QT_t5-50-500_sg} and Fig.~\ref{fig:cliff_QT_t5-50-500_r8_sg}, with $Q_T \approx 52$ for $t=500$.

For the $\mathcal{C}_{\mu, 8}$ ensemble, the additional Haar-random gate of order $r=8$ has two possible forms
\begin{equation}\label{eq:best_cliff_forms}
     U^\dagger P(\pi/4) U \quad \mathrm{and} \quad  U^\dagger P(3\pi/4) U,
\end{equation}
where $U$ is Haar-random. These two cases correspond to the rotation on the Bloch sphere by $\pi/4$ and $3\pi/4$ around a random axis. The best $T$-QCO upper bound found in our numerical computations is $Q_T \approx 3.7$ for $t=500$, which is close to the optimal value $\overline Q_\mathrm{opt} \approx 3.4$. It can be attained for the second form from \eqref{eq:best_cliff_forms} with $U$ being a Bloch sphere rotation around any axis $(x, y, 0)$ with $|x| \neq |y|$ by an angle in $[\pi/8, \pi/2]$. Interestingly, the worst $T$-QCO upper bound with $Q_T \approx 52$ for $t=500$ was achieved when $U$ was an element of the Clifford group.

\subsection{Hurwitz group}
The one-qubit Hurwitz subgroup $\mathcal{H} \subset \mathbf{U}(2)$ has 12 elements. We write
\begin{equation}
G_{\mathcal H}=\left\{
    \begin{pmatrix}
        i & 0 \\
        0 &-i
    \end{pmatrix},
    \begin{pmatrix}
        1 & 1 \\
        i & -i
    \end{pmatrix}
    \right\},
\end{equation}
so that $\mathcal{H}=\langle G_{\mathcal H}\rangle$, up to normalization. In the $T$-QCO computation, the Hurwitz adjoint orbit of the added gate is reduced to distinct elements of $\mathcal S_T$. The special choice of costly $T$ gate is the Super-Golden gate (of order $r=2$), denoted $T_{12}$

\begin{equation}
     T_{12}=
     \begin{pmatrix}
        3 & 1-i \\
        1+i & -3
    \end{pmatrix} \mathrm{,}
\end{equation}
up to normalization.
For the $\mathcal{H}_{\mu, 2}$ ensemble, the additional Haar-random gate of order $r=2$ is a Bloch sphere rotation by $\pi$ around a random axis. According to our numerical results, the optimal $T$-QCO bound $\overline Q_\mathrm{opt} \approx 4$ is attained for a Super-Golden gate set $\mathcal{H}_{T_{12}}$, where $T_{12}$ is a rotation around $(1, 1, \sqrt{9})/\sqrt{11}$. Computations for random gates also showed that the best $Q_T \approx4.1$ for $t=500$ is obtained for gates close to $T_{12}$.

\section{Conclusions and future directions}

In this paper, we introduce the Quantum Circuit Overhead (QCO) and the related notion of $T$-Quantum Circuit Overhead ($T$-QCO) as measures of the efficiency of universal quantum gate sets. QCO compares a finite gate set with optimal gate sets of the same size. The $T$-QCO applies the same idea after reducing a cheap/costly decomposition to the effective set $\mathcal S_T$ of gates obtainable with one counted operation. The concept of overhead can be applied to various NISQ and fault-tolerant architectures as a reasonable first approximation of the real cost-effectiveness of gate sets. We provide formulas for $Q$ and $Q_T$, which are the upper bounds on QCO and $T$-QCO, respectively, as well as their asymptotically optimal values (lower bounds) for all settings considered in the numerical examples. We performed extensive numerical calculations on a supercomputing cluster to study various random ensembles of universal single-qubit gate sets, particularly those derived as completions of the Clifford and Hurwitz groups with a Haar-random gate of infinite or finite order $r$. In our experiments, we compare various gate sets using the $Q/Q_T$ quantity.

Our numerical examples demonstrate that computing upper bounds on ($T$-)QCO is tractable on existing supercomputing infrastructure, at least for single-qubit gate sets, with the $Q/Q_T$ distributions stabilizing rapidly. Generic gate sets $\mathcal{S}_{\mu,n, r}$ consistently scored better in $Q/Q_T$ than the structured ones considered here. Interestingly, in the case of the Clifford group, the gate sets completed with the $P(\pi/4)$ gate turned out to perform significantly worse than the generic completions in terms of $Q_T$. Moreover, our analysis shows that the $P(\pi/4)$ gate is a highly non-optimal choice among the gates of order $r=8$ in this metric. In this case, we identified the best-performing gates of the same order as the family of the conjugates of $P(3\pi/4)$ by the Bloch sphere rotation around any axis $(x, y, 0)$ with $|x| \neq |y|$ by an angle in $[\pi/8, \pi/2]$. Finally, our results suggest that so-called single-qubit Super-Golden gates based on the Hurwitz group enjoy the optimal asymptotic value of $Q_T$. Interestingly, it does not seem to be the case for the Clifford group construction.

Clearly, one should be cautious about drawing conclusions about the overhead from the comparison of the upper bounds $Q/Q_T$. Preliminary finite-size tests of exact synthesis lengths are compatible with small values of $Q$ being associated with small overhead, but they do not establish that $Q$ or $Q_T$ tightly predicts exact overhead in general. The relation between these spectral upper bounds and exact synthesis cost therefore remains an important separate problem.

Moreover, the optimization of gates based on $T$-QCO is relevant in the quantum computing context only when the gates included in the $T$-count have comparable cost and the implementation overhead of elements of $C$ is negligible on that scale. If $C$ is generated by native cheap gates $G_C$, this means that $\operatorname{diam}(C,G_C)$ should be small compared with the relevant cost separation. For continuous cheap gate groups $C=\overline{\langle G_C\rangle}$, the same caveat applies to the corresponding $\epsilon$-diameter at the precision required by the circuit synthesis task. Strongly nonuniform costs among the gates included in the $T$-count, or a large cost for synthesizing elements of $C$, would require a refined cost model rather than the unweighted moment operator of the derived ensemble studied here.

As future directions, it would be interesting to perform numerical experiments for gate sets with larger locality, particularly those containing entangling gates. Such an approach may help identify good entangling gates within some parametrized families. Additionally, one would like to find and study fault-tolerant architectures that admit efficient implementations of the conjugate of $P(3\pi/4)$, identified here, to enhance the practical importance of this result. Finally, although the explicit calculation of ($T$-)QCO is, in general, intractable, it may be worthwhile to extend our preliminary analysis further to study smaller values of $\epsilon$.

\section*{Acknowledgments}
This research was funded by the National Science Centre, Poland under the grant OPUS: UMO2020/37/B/ST2/02478. We gratefully acknowledge Polish high-performance computing infrastructure PLGrid (HPC Center: ACK Cyfronet AGH) for providing computer facilities and support within computational grant no. PLG/2024/017436.

\section*{Data availability}

The code used in the numerical experiments is publicly available \cite{qco_code}.

\bibliography{pra_paper}

@ARTICLE{9614165,
  author={Oszmaniec, M. and Sawicki, A. and Horodecki, M.},
  journal={IEEE Transactions on Information Theory}, 
  title={Epsilon-Nets, Unitary Designs, and Random Quantum Circuits}, 
  year={2022},
  volume={68},
  number={2},
  pages={989-1015},
  doi={10.1109/TIT.2021.3128110}
}

@misc{kuperberg2023breakingcubicbarriersolovaykitaev,
      title={Breaking the cubic barrier in the {S}olovay-{K}itaev algorithm}, 
      author={G. Kuperberg},
      year={2023},
      eprint={2306.13158},
      archivePrefix={arXiv},
      primaryClass={quant-ph},
      url={https://arxiv.org/abs/2306.13158}, 
}

@misc{bouland2021efficientuniversalquantumcompilation,
      title={Efficient Universal Quantum Compilation: An Inverse-free {S}olovay-{K}itaev Algorithm}, 
      author={A.Bouland and T. Giurgica-Tiron},
      year={2021},
      eprint={2112.02040},
      archivePrefix={arXiv},
      primaryClass={quant-ph},
      url={https://arxiv.org/abs/2112.02040}, 
}

@misc{qco_code,
      howpublished="\url{https://github.com/pdulian/qco}" 
}

@article{S_owik_2023,
   title={Calculable lower bounds on the efficiency of universal sets of quantum gates},
   volume={56},
   ISSN={1751-8121},
   url={http://dx.doi.org/10.1088/1751-8121/acbd24},
   DOI={10.1088/1751-8121/acbd24},
   number={11},
   journal={Journal of Physics A: Mathematical and Theoretical},
   publisher={IOP Publishing},
   author={Słowik, O. and Sawicki, A.},
   year={2023},
   month=mar, pages={115304} }

@misc{słowik2025fundamentalsolutionsheatequation,
      title={Fundamental solutions of heat equation on unitary groups establish an improved relation between $\epsilon$-nets and approximate unitary $t$-designs}, 
      author={Oskar Słowik and Oliver Reardon-Smith and Adam Sawicki},
      year={2025},
      eprint={2503.08577},
      archivePrefix={arXiv},
      primaryClass={quant-ph},
      url={https://arxiv.org/abs/2503.08577}, 
}

@article{Harrow_2002,
   title={Efficient discrete approximations of quantum gates},
   volume={43},
   ISSN={1089-7658},
   url={http://dx.doi.org/10.1063/1.1495899},
   DOI={10.1063/1.1495899},
   number={9},
   journal={Journal of Mathematical Physics},
   publisher={AIP Publishing},
   author={Harrow, A. W. and Recht, B. and Chuang, I. L.},
   year={2002},
   month=sep, pages={4445–4451} }

@article{szarek1998,
    author = "S. J. Szarek",
    title = "Metric Entropy of Homogeneous Spaces",
    year = "1998",
    journal = "Banach Center Publications",
    volume="43"
}

@article{Bourgain2007OnTS,
  title={On the spectral gap for finitely-generated subgroups of $\mathrm{SU}(2)$
 
 
 
 
},
  author={J. Bourgain and A. Gamburd},
  journal={Inventiones mathematicae},
  year={2007},
  volume={171},
  pages={83-121},
  url={https://api.semanticscholar.org/CorpusID:122015109}
}

@misc{bourgain2011spectralgaptheoremsud,
      title={A Spectral Gap Theorem in $\mathrm{SU}(d)$}, 
      author={J. Bourgain and A. Gamburd},
      year={2011},
      eprint={1108.6264},
      archivePrefix={arXiv},
      primaryClass={math.GR},
      url={https://arxiv.org/abs/1108.6264}, 
}

@misc{benoist2014spectralgaptheoremsimple,
      title={A spectral gap theorem in simple {L}ie groups}, 
      author={Y. Benoist and N. de Saxcé},
      year={2014},
      eprint={1405.1808},
      archivePrefix={arXiv},
      primaryClass={math.RT},
      url={https://arxiv.org/abs/1405.1808}, 
}

@article{Dulian_2023,
   title={Matrix concentration inequalities and efficiency of random universal sets of quantum gates},
   volume={7},
   ISSN={2521-327X},
   url={http://dx.doi.org/10.22331/q-2023-04-20-983},
   DOI={10.22331/q-2023-04-20-983},
   journal={Quantum},
   publisher={Verein zur Förderung des Open Access Publizierens in den Quantenwissenschaften},
   author={Dulian, P. and Sawicki, A.},
   year={2023},
   month=apr, pages={983} }

@article{Parzanchevski_2018,
   title={Super-Golden-Gates for $\mathrm{PU}(2)$},
   volume={327},
   ISSN={0001-8708},
   url={http://dx.doi.org/10.1016/j.aim.2017.06.022},
   DOI={10.1016/j.aim.2017.06.022},
   journal={Advances in Mathematics},
   publisher={Elsevier BV},
   author={Parzanchevski, O. and Sarnak, P.},
   year={2018},
   month=mar, pages={869–901} }

@article{varju13,
    author = "P. P. Varj\'u",
    title = "Random walks in compact groups",
    year = "2013",
    journal = "Documenta Mathematica",
    volume= "18",
    pages = "1137--1175"
}

@article{kesten59,
    author = "H. Kesten",
    title = "Symmetric Random Walks on Groups",
    year = "1959",
    journal = "Transactions of the American Mathematical Society",
    volume = "92",
    issue = "2",
    pages = "336--354"
}

@article{lps86,
    author = "A. Lubotzky and R. Phillips and P. Sarnak",
    title = "Hecke operators and distributing points on the Sphere {I}",
    year = "1986",
    journal = "Communications on Pure and Applied Mathematics. Supplement: Proceedings of the Symposium on Frontiers of the Mathematical Sciences: 1985.",
    volume = "39",
    issue = "S1",
    pages = "S149-S186"
}

@article{lps87,
    author = "A. Lubotzky and R. Phillips and P. Sarnak",
    title = "Hecke operators and distributing points on {$S^2$}. {II}",
    year = "1987",
    journal = "Communications on Pure and Applied Mathematics",
    volume = "40",
    issue = "4",
    pages = "401-420"
}

@article{bocharov2013,
    author = "A. Bocharov and Y. Gurevich and K. M. Svore",
    title = "Efficient decomposition of single-qubit gates into {V} basis circuits",
    year = "2013",
    journal = "Physical Review A",
    volume = "88",
    issue = "1"
}

@article{kliuch2016,
    author = "V. Kliuchnikov and D. Maslov and M. Mosca",
    title = "Practical Approximation of Single-Qubit Unitaries by Single-Qubit Quantum {C}lifford and {T} Circuits",
    year = "2016",
    journal = "IEEE Transactions on Computers",
    volume = "65",
    pages = "161--172"
}

@article{selinger2015,
    author = "P. Selinger",
    title = "Efficient {C}lifford+{T} approximation of single-qubit operators",
    year = "2015",
    journal = "Quantum Information and Computation",
    volume = "15",
    issue = "1-2",
    pages = "159--180"
}

@misc{sarnak2015,
    author = "P. Sarnak",
    title = "Letter to {S}cott {A}aronson and {A}ndy {P}ollington on the {S}olovay-{K}itaev theorem",
    year = "2015"
}

@article{Dulian_2024,
  author={Dulian, Piotr and Sawicki, Adam},
  journal={IEEE Transactions on Information Theory}, 
  title={A Random Matrix Model for Random Approximate t-Designs}, 
  year={2024},
  volume={70},
  number={4},
  pages={2637-2654},
  doi={10.1109/TIT.2024.3367787}}

@book{barut,
    author = "Barut, Asim Orhan and R\k{a}czka, Ryszard",
    title = "Theory of group representations and applications",
    year = "1986",
    publisher = "World Scientific Publishing Co Pte Ltd.",
    isbn = "978-9971502164"
}

@article{benkart1994,
    author = "Benkart, Georgia and Chakrabarti, Mainsh and Halverson, Thomas and Leduc, Robert and Lee, Chanyoung and Stroomer, Jeffrey",
    title = "Tensor product representations of general linear groups and their connections with {B}rauer algebras",
    year = "1994",
    journal = "J. Algebra",
    volume = "166",
    pages = "529–567"
}

@article{PhysRevX.14.041068,
  title = {Saturation and Recurrence of Quantum Complexity in Random Local Quantum Dynamics},
  author = {Oszmaniec, Michał and Kotowski, Marcin and Horodecki, Michał and Hunter-Jones, Nicholas},
  journal = {Phys. Rev. X},
  volume = {14},
  issue = {4},
  pages = {041068},
  numpages = {43},
  year = {2024},
  month = {Dec},
  publisher = {American Physical Society},
  doi = {10.1103/PhysRevX.14.041068},
  url = {https://link.aps.org/doi/10.1103/PhysRevX.14.041068}
}

@book{Nielsen_Chuang_2010, place={Cambridge}, title={Quantum Computation and Quantum Information: 10th Anniversary Edition}, publisher={Cambridge University Press}, author={Nielsen, Michael A. and Chuang, Isaac L.}, year={2010}}

@book{Kitaev_Yu_Shen_2002,
author = {Kitaev, A. Yu. and Shen, A. H. and Vyalyi, M. N.},
title = {Classical and Quantum Computation},
year = {2002},
isbn = {0821832298},
publisher = {American Mathematical Society},
address = {USA}
}

@article{Gheorghiu_2023,
   title={Reducing the {CNOT} Count for {C}lifford+{T} Circuits on {NISQ} Architectures},
   volume={42},
   ISSN={1937-4151},
   url={http://dx.doi.org/10.1109/TCAD.2022.3213210},
   DOI={10.1109/tcad.2022.3213210},
   number={6},
   journal={IEEE Transactions on Computer-Aided Design of Integrated Circuits and Systems},
   publisher={Institute of Electrical and Electronics Engineers (IEEE)},
   author={Gheorghiu, Vlad and Huang, Jiaxin and Li, Sarah Meng and Mosca, Michele and Mukhopadhyay, Priyanka},
   year={2023},
   month=jun, pages={1873–1884} }

@article{Gheorghiu_2022,
   title={{T}-count and {T}-depth of any multi-qubit unitary},
   volume={8},
   ISSN={2056-6387},
   url={http://dx.doi.org/10.1038/s41534-022-00651-y},
   DOI={10.1038/s41534-022-00651-y},
   number={1},
   journal={npj Quantum Information},
   publisher={Springer Science and Business Media LLC},
   author={Gheorghiu, Vlad and Mosca, Michele and Mukhopadhyay, Priyanka},
   year={2022},
   month=nov }

@article{PhysRevLett.102.110502,
  title = {Restrictions on Transversal Encoded Quantum Gate Sets},
  author = {Eastin, Bryan and Knill, Emanuel},
  journal = {Phys. Rev. Lett.},
  volume = {102},
  issue = {11},
  pages = {110502},
  numpages = {4},
  year = {2009},
  month = {Mar},
  publisher = {American Physical Society},
  doi = {10.1103/PhysRevLett.102.110502},
  url = {https://link.aps.org/doi/10.1103/PhysRevLett.102.110502}
}

@article{Woods_2020,
   title={Continuous groups of transversal gates for quantum error correcting codes from finite clock reference frames},
   volume={4},
   ISSN={2521-327X},
   url={http://dx.doi.org/10.22331/q-2020-03-23-245},
   DOI={10.22331/q-2020-03-23-245},
   journal={Quantum},
   publisher={Verein zur Förderung des Open Access Publizierens in den Quantenwissenschaften},
   author={Woods, Mischa P. and Alhambra, {\'A}lvaro M.},
   year={2020},
   month=mar, pages={245} }

@article{Faist_2020,
   title={Continuous Symmetries and Approximate Quantum Error Correction},
   volume={10},
   ISSN={2160-3308},
   url={http://dx.doi.org/10.1103/PhysRevX.10.041018},
   DOI={10.1103/physrevx.10.041018},
   number={4},
   journal={Physical Review X},
   publisher={American Physical Society (APS)},
   author={Faist, Philippe and Nezami, Sepehr and Albert, Victor V. and Salton, Grant and Pastawski, Fernando and Hayden, Patrick and Preskill, John},
   year={2020},
   month=oct }

@article{Preskill2018quantumcomputingin,
  doi = {10.22331/q-2018-08-06-79},
  url = {https://doi.org/10.22331/q-2018-08-06-79},
  title = {Quantum {C}omputing in the {NISQ} era and beyond},
  author = {Preskill, John},
  journal = {{Quantum}},
  issn = {2521-327X},
  publisher = {{Verein zur Förderung des Open Access Publizierens in den Quantenwissenschaften}},
  volume = {2},
  pages = {79},
  month = aug,
  year = {2018}
}

@article{Horsman_2012,
   title={Surface code quantum computing by lattice surgery},
   volume={14},
   ISSN={1367-2630},
   url={http://dx.doi.org/10.1088/1367-2630/14/12/123011},
   DOI={10.1088/1367-2630/14/12/123011},
   number={12},
   journal={New Journal of Physics},
   publisher={IOP Publishing},
   author={Horsman, Dominic and Fowler, Austin G and Devitt, Simon and Meter, Rodney Van},
   year={2012},
   month=dec, pages={123011} }

@article{PhysRevA.76.012305,
  title = {Optimal resources for topological two-dimensional stabilizer codes: Comparative study},
  author = {Bombin, H. and Martin-Delgado, M. A.},
  journal = {Phys. Rev. A},
  volume = {76},
  issue = {1},
  pages = {012305},
  numpages = {6},
  year = {2007},
  month = {Jul},
  publisher = {American Physical Society},
  doi = {10.1103/PhysRevA.76.012305},
  url = {https://link.aps.org/doi/10.1103/PhysRevA.76.012305}
}

@article{PhysRevA.91.032330,
  title = {Universal transversal gates with color codes: A simplified approach},
  author = {Kubica, Aleksander and Beverland, Michael E.},
  journal = {Phys. Rev. A},
  volume = {91},
  issue = {3},
  pages = {032330},
  numpages = {12},
  year = {2015},
  month = {Mar},
  publisher = {American Physical Society},
  doi = {10.1103/PhysRevA.91.032330},
  url = {https://link.aps.org/doi/10.1103/PhysRevA.91.032330}
}

@article{PhysRevA.86.052329,
  title = {Magic-state distillation with low overhead},
  author = {Bravyi, Sergey and Haah, Jeongwan},
  journal = {Phys. Rev. A},
  volume = {86},
  issue = {5},
  pages = {052329},
  numpages = {10},
  year = {2012},
  month = {Nov},
  publisher = {American Physical Society},
  doi = {10.1103/PhysRevA.86.052329},
  url = {https://link.aps.org/doi/10.1103/PhysRevA.86.052329}
}

@article{RevModPhys.80.1083,
  title = {Non-Abelian anyons and topological quantum computation},
  author = {Nayak, Chetan and Simon, Steven H. and Stern, Ady and Freedman, Michael and Das Sarma, Sankar},
  journal = {Rev. Mod. Phys.},
  volume = {80},
  issue = {3},
  pages = {1083--1159},
  numpages = {0},
  year = {2008},
  month = {Sep},
  publisher = {American Physical Society},
  doi = {10.1103/RevModPhys.80.1083},
  url = {https://link.aps.org/doi/10.1103/RevModPhys.80.1083}
}

@article{PhysRevLett.104.180505,
  title = {Implementing Arbitrary Phase Gates with Ising Anyons},
  author = {Bonderson, Parsa and Clarke, David J. and Nayak, Chetan and Shtengel, Kirill},
  journal = {Phys. Rev. Lett.},
  volume = {104},
  issue = {18},
  pages = {180505},
  numpages = {4},
  year = {2010},
  month = {May},
  publisher = {American Physical Society},
  doi = {10.1103/PhysRevLett.104.180505},
  url = {https://link.aps.org/doi/10.1103/PhysRevLett.104.180505}
}

@misc{bombin2015gaugecolorcodesoptimal,
      title={Gauge Color Codes: Optimal Transversal Gates and Gauge Fixing in Topological Stabilizer Codes}, 
      author={H. Bombin},
      year={2015},
      eprint={1311.0879},
      archivePrefix={arXiv},
      primaryClass={quant-ph},
      url={https://arxiv.org/abs/1311.0879}, 
}

@article{PhysRevA.100.012312,
  title = {Three-dimensional surface codes: Transversal gates and fault-tolerant architectures},
  author = {Vasmer, Michael and Browne, Dan E.},
  journal = {Phys. Rev. A},
  volume = {100},
  issue = {1},
  pages = {012312},
  numpages = {20},
  year = {2019},
  month = {Jul},
  publisher = {American Physical Society},
  doi = {10.1103/PhysRevA.100.012312},
  url = {https://link.aps.org/doi/10.1103/PhysRevA.100.012312}
}

@article{Litinski_2019,
   title={A Game of Surface Codes: Large-Scale Quantum Computing with Lattice Surgery},
   volume={3},
   ISSN={2521-327X},
   url={http://dx.doi.org/10.22331/q-2019-03-05-128},
   DOI={10.22331/q-2019-03-05-128},
   journal={Quantum},
   publisher={Verein zur Forderung des Open Access Publizierens in den Quantenwissenschaften},
   author={Litinski, Daniel},
   year={2019},
   month=mar, pages={128} }

@article{Noh_2020,
   title={Efficient classical simulation of noisy random quantum circuits in one dimension},
   volume={4},
   ISSN={2521-327X},
   url={http://dx.doi.org/10.22331/q-2020-09-11-318},
   DOI={10.22331/q-2020-09-11-318},
   journal={Quantum},
   publisher={Verein zur Förderung des Open Access Publizierens in den Quantenwissenschaften},
   author={Noh, Kyungjoo and Jiang, Liang and Fefferman, Bill},
   year={2020},
   month=sep, pages={318} }

@misc{gottesman2005quantumerrorcorrectionfaulttolerance,
      title={Quantum Error Correction and Fault-Tolerance}, 
      author={Daniel Gottesman},
      year={2005},
      eprint={quant-ph/0507174},
      archivePrefix={arXiv},
      primaryClass={quant-ph},
      url={https://arxiv.org/abs/quant-ph/0507174}, 
}

@misc{zhou2024algorithmicfaulttolerancefast,
      title={Algorithmic Fault Tolerance for Fast Quantum Computing}, 
      author={Hengyun Zhou and Chen Zhao and Madelyn Cain and Dolev Bluvstein and Casey Duckering and Hong-Ye Hu and Sheng-Tao Wang and Aleksander Kubica and Mikhail D. Lukin},
      year={2024},
      eprint={2406.17653},
      archivePrefix={arXiv},
      primaryClass={quant-ph},
      url={https://arxiv.org/abs/2406.17653}, 
}

@misc{ge2024quantumcircuitsynthesiscompilation,
      title={Quantum Circuit Synthesis and Compilation Optimization: Overview and Prospects}, 
      author={Yan Ge and Wu Wenjie and Chen Yuheng and Pan Kaisen and Lu Xudong and Zhou Zixiang and Wang Yuhan and Wang Ruocheng and Yan Junchi},
      year={2024},
      eprint={2407.00736},
      archivePrefix={arXiv},
      primaryClass={quant-ph},
      url={https://arxiv.org/abs/2407.00736}, 
}

@Article{Ruiz2025,
author={Ruiz, Francisco J. R.
and Laakkonen, Tuomas
and Bausch, Johannes
and Balog, Matej
and Barekatain, Mohammadamin
and Heras, Francisco J. H.
and Novikov, Alexander
and Fitzpatrick, Nathan
and Romera-Paredes, Bernardino
and van de Wetering, John
and Fawzi, Alhussein
and Meichanetzidis, Konstantinos
and Kohli, Pushmeet},
title={Quantum circuit optimization with {A}lpha{T}ensor},
journal={Nature Machine Intelligence},
year={2025},
month={Mar},
day={01},
volume={7},
number={3},
pages={374-385},
issn={2522-5839},
doi={10.1038/s42256-025-01001-1},
url={https://doi.org/10.1038/s42256-025-01001-1}
}

@misc{vandaele2024lowertcountfasteralgorithms,
      title={Lower {$T$}-count with faster algorithms}, 
      author={Vivien Vandaele},
      year={2024},
      eprint={2407.08695},
      archivePrefix={arXiv},
      primaryClass={quant-ph},
      url={https://arxiv.org/abs/2407.08695}, 
}

@article{10.5555/2685179.2685180,
author = {Gosset, David and Kliuchnikov, Vadym and Mosca, Michele and Russo, Vincent},
title = {An algorithm for the {T}-count},
year = {2014},
issue_date = {November 2014},
publisher = {Rinton Press, Incorporated},
address = {Paramus, NJ},
volume = {14},
number = {15–16},
issn = {1533-7146},
journal = {Quantum Info. Comput.},
month = nov,
pages = {1261–1276},
numpages = {16}
}

@article{
doi:10.1073/pnas.2026250118,
author = {Yi-Han Luo  and Ming-Cheng Chen  and Manuel Erhard  and Han-Sen Zhong  and Dian Wu  and Hao-Yang Tang  and Qi Zhao  and Xi-Lin Wang  and Keisuke Fujii  and Li Li  and Nai-Le Liu  and Kae Nemoto  and William J. Munro  and Chao-Yang Lu  and Anton Zeilinger  and Jian-Wei Pan },
title = {Quantum teleportation of physical qubits into logical code spaces},
journal = {Proceedings of the National Academy of Sciences},
volume = {118},
number = {36},
pages = {e2026250118},
year = {2021},
doi = {10.1073/pnas.2026250118},
URL = {https://www.pnas.org/doi/abs/10.1073/pnas.2026250118},
eprint = {https://www.pnas.org/doi/pdf/10.1073/pnas.2026250118}}

@article{Eastin_2009,
   title={Restrictions on Transversal Encoded Quantum Gate Sets},
   volume={102},
   ISSN={1079-7114},
   url={http://dx.doi.org/10.1103/PhysRevLett.102.110502},
   DOI={10.1103/physrevlett.102.110502},
   number={11},
   journal={Physical Review Letters},
   publisher={American Physical Society (APS)},
   author={Eastin, Bryan and Knill, Emanuel},
   year={2009},
   month=mar }

@article{Tokusumi_2018,
   title={Quantum circuit model of black hole evaporation},
   volume={35},
   ISSN={1361-6382},
   url={http://dx.doi.org/10.1088/1361-6382/aaeb5a},
   DOI={10.1088/1361-6382/aaeb5a},
   number={23},
   journal={Classical and Quantum Gravity},
   publisher={IOP Publishing},
   author={Tokusumi, Tomoro and Matsumura, Akira and Nambu, Yasusada},
   year={2018},
   month=nov, pages={235013} }

@article{Fisher_2023,
   title={Random Quantum Circuits},
   volume={14},
   ISSN={1947-5462},
   url={http://dx.doi.org/10.1146/annurev-conmatphys-031720-030658},
   DOI={10.1146/annurev-conmatphys-031720-030658},
   number={1},
   journal={Annual Review of Condensed Matter Physics},
   publisher={Annual Reviews},
   author={Fisher, Matthew P.A. and Khemani, Vedika and Nahum, Adam and Vijay, Sagar},
   year={2023},
   month=mar, pages={335–379} }

@article{Hayden_2007,
   title={Black holes as mirrors: quantum information in random subsystems},
   volume={2007},
   ISSN={1029-8479},
   url={http://dx.doi.org/10.1088/1126-6708/2007/09/120},
   DOI={10.1088/1126-6708/2007/09/120},
   number={09},
   journal={Journal of High Energy Physics},
   publisher={Springer Science and Business Media LLC},
   author={Hayden, Patrick and Preskill, John},
   year={2007},
   month=sep, pages={120–120} }

@article{PRXQuantum.2.020341,
  title = {Cost of Universality: A Comparative Study of the Overhead of State Distillation and Code Switching with Color Codes},
  author = {Beverland, Michael E. and Kubica, Aleksander and Svore, Krysta M.},
  journal = {PRX Quantum},
  volume = {2},
  issue = {2},
  pages = {020341},
  numpages = {46},
  year = {2021},
  month = {Jun},
  publisher = {American Physical Society},
  doi = {10.1103/PRXQuantum.2.020341},
  url = {https://link.aps.org/doi/10.1103/PRXQuantum.2.020341}
}

@misc{dawson2005solovaykitaevalgorithm,
      title={The {S}olovay-{K}itaev algorithm}, 
      author={Christopher M. Dawson and Michael A. Nielsen},
      year={2005},
      eprint={quant-ph/0505030},
      archivePrefix={arXiv},
      primaryClass={quant-ph},
      url={https://arxiv.org/abs/quant-ph/0505030}, 
}

@article{H_ner_2018,
   title={A software methodology for compiling quantum programs},
   volume={3},
   ISSN={2058-9565},
   url={http://dx.doi.org/10.1088/2058-9565/aaa5cc},
   DOI={10.1088/2058-9565/aaa5cc},
   number={2},
   journal={Quantum Science and Technology},
   publisher={IOP Publishing},
   author={Häner, Thomas and Steiger, Damian S and Svore, Krysta and Troyer, Matthias},
   year={2018},
   month=feb, pages={020501} }

@misc{heyfron2018efficientquantumcompilerreduces,
      title={An Efficient Quantum Compiler that reduces {$T$} count}, 
      author={Luke Heyfron and Earl T. Campbell},
      year={2018},
      eprint={1712.01557},
      archivePrefix={arXiv},
      primaryClass={quant-ph},
      url={https://arxiv.org/abs/1712.01557}, 
}

@article{Claeys_2022,
   title={Exact dynamics in dual-unitary quantum circuits with projective measurements},
   volume={4},
   ISSN={2643-1564},
   url={http://dx.doi.org/10.1103/PhysRevResearch.4.043212},
   DOI={10.1103/physrevresearch.4.043212},
   number={4},
   journal={Physical Review Research},
   publisher={American Physical Society (APS)},
   author={Claeys, Pieter W. and Henry, Marius and Vicary, Jamie and Lamacraft, Austen},
   year={2022},
   month=dec }

@misc{aaronson2016complexityquantumstatestransformations,
      title={The Complexity of Quantum States and Transformations: From Quantum Money to Black Holes}, 
      author={Scott Aaronson},
      year={2016},
      eprint={1607.05256},
      archivePrefix={arXiv},
      primaryClass={quant-ph},
      url={https://arxiv.org/abs/1607.05256}, 
}

@Article{Dolgopyat2002,
author={Dolgopyat, Dmitry},
title={On mixing properties of compact group extensions of hyperbolic systems},
journal={Israel Journal of Mathematics},
year={2002},
month={Dec},
day={01},
volume={130},
number={1},
pages={157-205},
issn={1565-8511},
doi={10.1007/BF02764076},
url={https://doi.org/10.1007/BF02764076}
}

@article{Fowler_2009,
   title={High-threshold universal quantum computation on the surface code},
   volume={80},
   ISSN={1094-1622},
   url={http://dx.doi.org/10.1103/PhysRevA.80.052312},
   DOI={10.1103/physreva.80.052312},
   number={5},
   journal={Physical Review A},
   publisher={American Physical Society (APS)},
   author={Fowler, Austin G. and Stephens, Ashley M. and Groszkowski, Peter},
   year={2009},
   month=nov }

\appendix

\section{Unitary channels and the projective group}
\label{app:unitary}
The unitary channel $\mathbf{U}$ acting on a Hilbert space $\mathcal{H} \cong \mathbb{C}^d$ is the CPTP map defined via $\mathbf{U}(\rho) = U \rho U^{\dagger}$, for any quantum state $\rho: \mathcal{H} \rightarrow \mathcal{H}$ and some fixed unitary representative $U$ from $\mathrm{U}(d)$. Since two unitaries $U, V$ that differ by a phase $U=Ve^{i \phi}$ define the same unitary channel, the group of all unitary channels $\mathbf{U}(d)$ can be identified with the projective unitary group $\mathrm{PU}(d)=\mathrm{U}(d)/U(1)$, where the canonical projection $\pi: \mathrm{U}(d) \rightarrow \mathbf{U}(d)$ maps unitaries to their corresponding unitary channels $U \mapsto \mathbf{U}$.

In practice, one is often interested in the closeness of different unitary channels. Various norms (and induced metrics) can be used to quantify it. A prominent example is the diamond norm $||\cdot||_{\diamond}$ and the induced metric
$d_{\diamond}\left(\mathbf{U},\mathbf{V}\right)=||\mathbf{U}-\mathbf{V}||_{\diamond}$. The diamond metric has a clear operational meaning in terms of the statistical distinguishability of two channels. The relationship between $d_{\diamond}$ and our metric $d$ \eqref{eq:d_m} is given by $d(\mathbf{U},\mathbf{V}) \leq d_{\diamond}(\mathbf{U},\mathbf{V}) \leq 2 \, d(\mathbf{U},\mathbf{V})$ \cite{9614165}.

\section{Approximate \texorpdfstring{$t$}{t}-designs and \texorpdfstring{$\epsilon$}{epsilon}-nets}
\label{app:t_designs}

Balanced polynomials of degree $t$ are homogeneous polynomials of degree $t$ in the matrix elements $u_{i,j}$ and degree $t$ in their complex conjugates $\overline{u}_{i,j}$. Such polynomials are well-defined on $\mathbf{U}(d)$ because they are insensitive to global phases. We denote the space of all such polynomials by $\mathcal{H}_t$. The space $\mathcal{H}_t$ is spanned by the entries of $U^{t,t}:=U^{\otimes t}\otimes \bar{U}^{\otimes t}$; therefore, each polynomial $f_t(U) \in \mathcal{H}_t$ can be expressed as
\begin{equation}
    f_t(U)=\mathrm{Tr}\left(A \left(U^{\otimes t}\otimes \bar{U}^{\otimes t}\right)\right)
\end{equation}
for some matrix $A$. Let $\mu$ be the normalized Haar measure on $\mathbf{U}(d)$, so that $\mu(\mathbf{U}(d))=1$. The Haar measure provides the natural notion of a uniform distribution on $\mathbf{U}(d)$.

For completeness, we spell out the exact-design formulation that underlies the approximate-design notation introduced in the main text. A $t$-design is a probability measure $\nu$ on $\mathbf{U}(d)$ that yields the same averages as the Haar measure for all polynomials $f_t(U) \in \mathcal{H}_t$:
\begin{equation}
    \int_{\mathbf{U}(d)}d\nu(U)f_t(U)=\int_{\mathbf{U}(d)}d\mu(U)f_t(U) \mathrm{.}
    \label{eq:tde_app}
\end{equation}

The case in which $\nu$ is supported on finitely many weighted points $\{(\nu_i,U_i)\}$ is of particular practical importance. In this case, Eq.~\eqref{eq:tde_app} becomes
\begin{equation}
\sum_{U_i\in\mathcal{S}}\nu_i f_t(U_i)=\int_{\mathbf{U}(d)} d\mu(U)f_t(U) \mathrm{,}
\label{eq:td_app}
\end{equation}
where $\mathcal{S}$ denotes the finite set supporting the measure $\nu$. We are mostly interested in uniform designs, for which all $\nu_i = 1/|\mathcal{S}|$; the corresponding measure is denoted by $\nu_{\mathcal{S}}$. Thus, when we say that a finite set $\mathcal{S} \subset \mathbf{U}(d)$ is a $t$-design, we mean that the uniform measure $\nu_{\mathcal{S}}$ is a $t$-design.

Using the moment operators of Eq.~\eqref{eq:t_moment}, the deviation from an exact $t$-design is measured by the discrepancy $\delta(\nu,t)$ defined in the main text. Hence, $\delta(\nu,t)=0$ corresponds to an exact $t$-design, whereas $\delta(\nu,t)<1$ gives the approximate-design notion used throughout the paper.

The uniform measure corresponding to all circuits of length $\ell$ is the convolution $\nu_{\mathcal{S}}^{\ast (\ell)}$, and its averaging operator is $T_{\nu_{\mathcal{S}}^{\ast (\ell)},t}=T_{\nu_{\mathcal{S}},t}^{\ell}$. Therefore,
\begin{equation}
\label{eq:ineq_app}
\left\|T^{\ell}_{\nu_{\mathcal{S}},t}-T_{\mu,t}\right\|_{\infty} \leq \delta^{\ell}(\nu_{\mathcal{S}},t) \mathrm{,}
\end{equation}
with equality for symmetric $\mathcal{S}$, for which the averaging operators are self-adjoint. This is the precise form of the circuit-averaging statement used in the main text.

The support of $\nu_{\mathcal{S}}^{\ast (\ell)}$ is an $\epsilon$-net if $\nu_{\mathcal{S}}^{\ast (\ell)}$ is a $\delta$-approximate $t$-design with $t(\epsilon)\gtrsim d^{5/2}/\epsilon$ and $\delta(\epsilon) \lesssim (\epsilon/d^{1/2})^{d^2}$; see Refs.~\cite{9614165, słowik2025fundamentalsolutionsheatequation} for precise formulas.

\section{Optimal spectral gap and Kesten--McKay measure}
\label{app:kesten}
 We discuss the applicability of the optimal value (\ref{eq:efficient_gates}) and the related measure in various settings considered in this paper.

For a symmetric (i.e., inverse-closed) gate set $\mathcal{S}$, the $t$-moment operator (\ref{eq:t_moment}) is a bounded self-adjoint operator with a well-defined spectrum. Its spectral measure $\sigma_{\mathcal{S},t}$ is compactly supported and hence is determined by its moments $\sigma_{\mathcal{S},t}^{(m)}$.
The asymptotic behavior of such moments, i.e., the limit $\lim_{t \to \infty}\sigma_{\mathcal{S},t}^{(m)}$ is determined by the number of length $m$ spellings of identity and was provided in \cite{kesten59} in the case of $\mathcal{S}$ generating a free group. Moreover, it was shown in \cite{kesten59} that in this case there exists a measure $\sigma_{\mathcal{S}}$ such that $\sigma_{\mathcal{S}}^{(m)}=\lim_{t \to \infty}\sigma_{\mathcal{S},t}^{(m)}$, known as the Kesten--McKay or Plancherel measure

 \begin{equation}
     d \sigma_{\mathcal{S}}(x) = \frac{|\mathcal{S}| \sqrt{\delta^2_{\mathrm{opt}}(\mathcal{S})-x^2}}{2 \pi (1-x^2)} \mathbf{1}_{[-\delta_{\mathrm{opt}}(\mathcal{S}),\, \delta_{\mathrm{opt}}(\mathcal{S})]} dx \mathrm{,}
 \end{equation}
where $\delta_{\mathrm{opt}}(\mathcal{S})$ is the optimal value (\ref{eq:efficient_gates}). This implies that $\sigma_{\mathcal{S},t}$ converges weakly to $\sigma_{\mathcal{S}}$ in the limit $t \to \infty$ (see \cite{Dulian_2024} for details).
Furthermore, analogous results can be obtained for any (i.e., not necessarily inverse-closed) finite $\mathcal{S}$, for which $\mathcal{S} \cup \mathcal{S}^{-1}$ generates a free group \cite{Dulian_2024}. However, since in this setting the $t$-moment operator need not be self-adjoint, by the Kesten--McKay measure we mean the spectral measure of $\sqrt{T_{\nu_{\mathcal{S}},t} T_{\nu_{\mathcal{S}},t}^*}$ as $t \to \infty$, or equivalently the measure describing the singular values of $T_{\nu_{\mathcal{S}},t}$ as $t \to \infty$, given by
\begin{equation}
    \frac{|\mathcal{S}| \sqrt{\delta^2_{\mathrm{opt}}(\mathcal{S})-x^2}}{\pi (1-x^2)} \mathbf{1}_{[0,\, \delta_{\mathrm{opt}}(\mathcal{S})]} dx \mathrm{.}
 \end{equation}

Thus, the Kesten--McKay measure can be applied in the setting of Haar-random gate sets $\mathcal{S}$, since then $\mathcal{S} \cup \mathcal{S}^{-1}$ generates a free group with probability 1.

\begin{figure}
  \centering
      \includegraphics[width=0.5\textwidth]{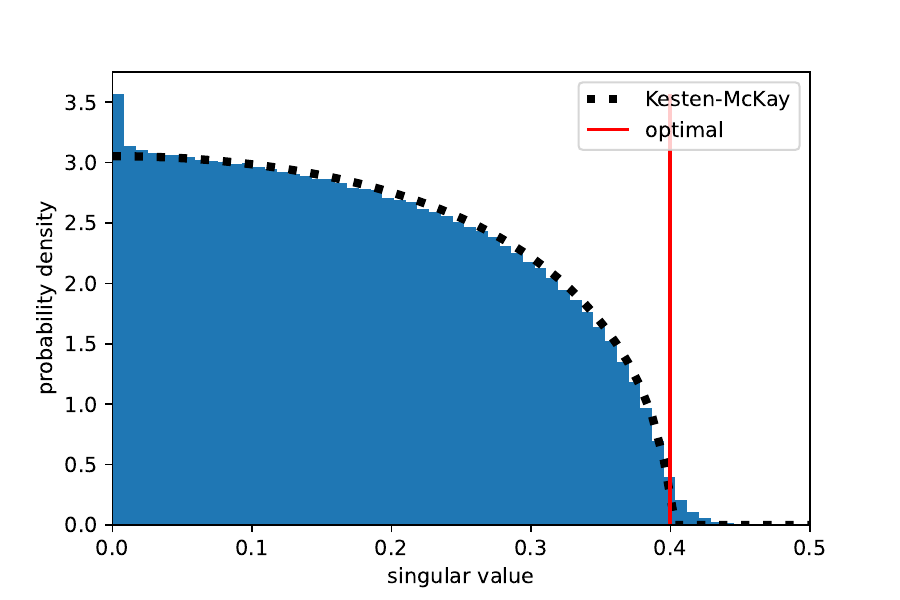}
  \caption{The probability density of the singular values of the $t$-moment operator for a derived ensemble of type $\mathcal{C}_{\mu,8}$ with approximately $20$ gate sets for $t=500$. The dotted line denotes the Kesten--McKay measure and the solid line denotes the corresponding optimal value.}
  \label{fig:cliff_t500_r8_spec}
\end{figure}

\begin{figure}
  \centering
      \includegraphics[width=0.5\textwidth]{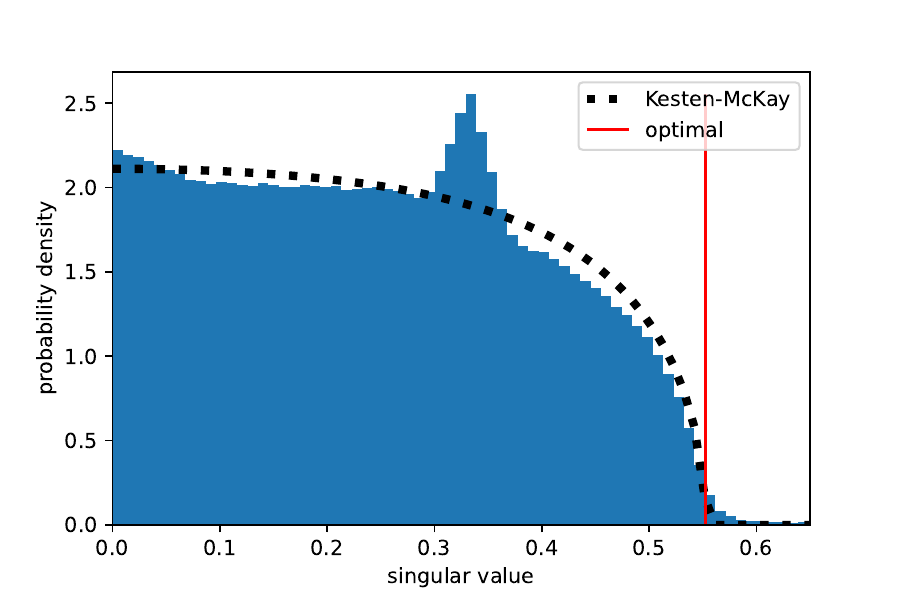}
  \caption{The probability density of the singular values of the $t$-moment operator for a derived ensemble of type $\mathcal{H}_{\mu,2}$ with approximately $20$ gate sets for $t=500$. The dotted line denotes the Kesten--McKay measure and the solid line denotes the corresponding optimal value.}
  \label{fig:hurwitz_t500_r2_spec}
\end{figure}

Crucially, the Kesten--McKay measure can also be applied in the $T$-QCO setting (\ref{eq:S_ansatz}) when the additional gate $T$ has infinite order (e.g., when it is Haar-random). This follows from the fact that in this case the derived gate set construction (\ref{eq:derived}), which is used to upper bound the $T$-QCO (\ref{eq:Q_T}), does not change the number of spellings of identity, compared with the free-group case. For a Haar-random gate $T$ of fixed finite order, the number of spellings of identity is increased, which implies that the (even) spectral measure moments are larger than the moments of the Kesten--McKay measure. As a consequence, the support of the Kesten--McKay measure is contained in the support of such a spectral measure and the bound (\ref{eq:efficient_gates}) can be applied. However, it was not clear how tight such a bound is with respect to the actual cutoff of the bulk spectrum. To verify it, we checked the distribution of the singular values of the $t$-moment operators for (derived) ensembles of type $C_{\mu, r}$ with finite $r$. The resulting distributions are close to the Kesten--McKay distribution, and the support of the latter is tightly contained in that of the former (see Figs.~\ref{fig:cliff_t500_r8_spec} and~\ref{fig:hurwitz_t500_r2_spec}).
Thus, the optimal value (\ref{eq:efficient_gates}) is relevant in all cases considered in this paper.

\section{Derived gate sets and the \texorpdfstring{$T$}{T}-QCO upper bound}
\label{app:T_QCO}

\subsection{Cheap gate group and diameter}

The native operations whose cost is small compared with the costly gates need not themselves be closed under composition. We denote such a native cheap generating set by $G_C$ and use the closed group
\begin{equation}
C=\overline{\langle G_C\rangle}\mathrm{.}
\end{equation}
For a finite generated group, the closure is redundant. The $T$-QCO model then neglects the cost of elements of $C$. This is an approximation: it is appropriate only when elements of $C$ can be implemented with cost small compared with one use of a costly gate $T_i$.

For finite $C$, this condition can be expressed using the diameter of $C$ with respect to $G_C$. Define the word length
\begin{equation}
 \ell_{G_C}(c)=\min\{m:\; c=g_1\cdots g_m,\; g_j\in G_C\cup G_C^{-1}\}\mathrm{.}
\end{equation}
Then
\begin{equation}
 \operatorname{diam}(C,G_C)=\max_{c\in C}\ell_{G_C}(c)
\end{equation}
is the diameter of the Cayley graph of $C$ generated by $G_C$. Thus every element of $C$ can be synthesized using at most $\operatorname{diam}(C,G_C)$ native cheap gates. Neglecting the cost of $C$ is justified when this worst-case implementation cost is negligible compared with the cost of one $T_i$. If the costs inside either class vary strongly, this unweighted counting model should be regarded only as a rough proxy; a more detailed weighted model would then be needed.

For continuous $C$, exact finite synthesis of all elements is usually impossible. The analogous notion is an $\epsilon$-diameter: $\operatorname{diam}_{\epsilon}(C,G_C)$ is the smallest word length $m$ for which words in $G_C\cup G_C^{-1}$ of length at most $m$ form an $\epsilon$-net of $C$ in the metric used in the main text. This is the same $\epsilon$-net idea used in the definition of circuit overhead: finite diameter gives exact coverage of a finite group, whereas finite $\epsilon$-diameter gives approximate coverage of a continuous group at resolution $\epsilon$. If a circuit of $T$-count $L$ must approximate a target to accuracy $\epsilon$, then the inserted elements of $C$ generally have to be synthesized at a finer scale, for example of order $\epsilon/L$ by a triangle-inequality estimate. If this overhead is not negligible, it should be included in the cost model rather than absorbed into $C$.

\subsection{Reduced derived set and symmetries}

The set $\mathcal S_T$ in Eq.~\eqref{eq:derived} is the union of the adjoint orbits of the gates $T_i$ under $C$. We write
\begin{equation}
C\cdot T_i:=\{cT_i c^\dagger:\;c\in C\}
\end{equation}
for the adjoint orbit of $T_i$, so that
\begin{equation}
\mathcal S_T=\bigcup_{i=1}^k C\cdot T_i\mathrm{.}
\end{equation}
It is important that $\mathcal S_T$ is the reduced set of distinct gates. If two pairs $(i,c)$ and $(j,c')$ give the same unitary, they give the same element of the derived gate set and should be counted only once in the volumetric comparison. This is the same convention as in ordinary QCO, where repeated gates do not change the gate set.

The inequality $|\mathcal S_T|\leq k|C|$ follows immediately from the definition. Strict inequality has two symmetry sources. First, a gate $T_i$ may have a nontrivial stabilizer in $C$,
\begin{equation}
\operatorname{Stab}_C(T_i)=\{c\in C:\; cT_i c^\dagger=T_i\}\mathrm{,}
\end{equation}
so that different cheap dressings produce the same derived gate. Second, two gates may belong to the same adjoint orbit, $T_j=cT_i c^\dagger$ for some $c\in C$. In that case one of them is redundant up to cheap gates. These repetitions should be removed before applying the volumetric bound: the true $T$-complexity is unchanged, while the reduced derived set has smaller support.

This also gives the physical meaning of the symmetry language. In NISQ examples, $C$ may be generated by cheap local controls and the adjoint orbits describe the different ways in which local controls dress an entangling gate. A stabilizer is a local control symmetry that leaves the entangling gate unchanged. In fault-tolerant or QEC examples, $C$ may be a Clifford group generated by cheap Clifford operations. The Clifford adjoint action can change the Pauli frame or orientation of a non-Clifford gate, while the Clifford stabilizer consists of Clifford operations that leave that non-Clifford gate unchanged.

\subsection{Reduction to an ordinary QCO bound}

We now justify Eq.~\eqref{eq:Q_T}. The main point is that a circuit over $\mathcal S=C\cup\{T_1,\ldots,T_k\}$ with a given $T$-count can be rewritten, up to initial or final multiplication by elements of $C$, as a word of the same length over the reduced derived set $\mathcal S_T$. Conversely, any word over $\mathcal S_T$ can be implemented with the same $T$-count together with cheap gates from $C$.

To see this, consider a $T$-count-minimal $\epsilon$-approximation of $U$ over $\mathcal S$. For $k=1$, and omitting inessential endpoints for readability, it can be written as
\begin{equation}
\label{eq:rwT}
  U \approx_{\epsilon} c_{i_1} T^{k_1} c_{i_2} T^{k_2} \ldots c_{i_p} T^{k_p} \mathrm{,}
\end{equation}
where $c_{i_j}\in C$, $k_j\in\mathbb N$, and the total $T$-count is
\begin{equation}
  L=\sum_{j=1}^p k_j \mathrm{.}
\end{equation}
Since $C$ is a group, the partial products $d_j=c_{i_1}c_{i_2}\cdots c_{i_j}$ also belong to $C$. Hence, with $g_j=d_jTd_j^\dagger\in\mathcal S_T$, the same approximation can be written as
\begin{equation}
  U \approx_{\epsilon} g_1^{k_1}g_2^{k_2}\cdots g_p^{k_p}d_p\mathrm{.}
\end{equation}
Thus it is a word of length $L$ in derived gates, followed by a final uncounted element of $C$. The general case with several $T_i$ and with endpoints present is analogous.

Therefore, to upper bound the supremal $T$-complexity of $\mathcal S$, it suffices to apply the ordinary circuit-complexity bound to the reduced derived set $\mathcal S_T$, with the endpoint gates from $C$ left uncounted. For finite $\mathcal S_T$, the same volumetric and spectral argument as in the QCO section gives
\begin{equation}
   \frac{\ell(\mathcal{S}_T,\epsilon)}{\ell_{\mathrm{opt}}(|\mathcal{S}_T|,\epsilon)}  \lesssim Q_T(\mathcal{S},\epsilon) \mathrm{,}
\end{equation}
with
\begin{equation}
     Q_T(\mathcal{S},\epsilon) = \frac{\log|\mathcal{S}_T|}{\log\left(1/\delta(\nu_{\mathcal{S}_T},t(\epsilon))\right)} \mathrm{.}
\end{equation}
Thus the original definition $Q_T(\mathcal S,\epsilon)=Q(\mathcal S_T,\epsilon)$ is the appropriate upper-bound proxy: it uses the reduced effective support, not the unreduced list of possible dressings.

\section{Numerical experiments---methods}
\label{app:num_expl}

To obtain the value of $Q(\mathcal{S}, \epsilon)$, one needs to compute the norm $\delta(\nu_\mathcal{S},t) =\left\|T_{\nu_\mathcal{S},t}-T_{\mu,t}\right\|_{\infty}$ (see Eq.~\eqref{eq:t_moment}). In a naive approach, one could compute $U^{t, t} = U^{\otimes t}\otimes\bar{U}^{\otimes t}$ for each $U$ in $\mathcal{S}$, but performing such a calculation is exponentially hard in $t$.
This problem can be avoided by noticing that the mapping $U \mapsto U^{t,t}$ is a representation of the $SU(d)$ group on a $d^{2t}$-dimensional space. Every representation of $SU(d)$ can be expressed as a block diagonal matrix, where each block is some irreducible representation (irrep) of $SU(d)$ \cite{barut}. In our case, it reads
\begin{gather}
    U^{t,t} = \blockdiagonal{\pi_{\lambda_1}(U)}{\pi_{\lambda_2}(U)}{\pi_{\lambda_k}(U)},
\end{gather}
where $\pi_\lambda$ is an irrep with label $\lambda$. It follows that the $t$-moment operators are block diagonal as well, and their blocks are given by $T_{\nu, \lambda} = \int_G d\nu(U) \pi_\lambda(U)$. Furthermore, by the orthogonality of irreps \cite{barut}, the Haar measure blocks $T_{\mu, \lambda}$ are equal to zero for all irreps $\pi_\lambda$, except the trivial one $\pi_0(U)=1$. In summary, the value of $\delta(\nu_\mathcal{S}, t)$ can be computed as
\begin{gather}
     \max_\lambda \|T_{\nu_\mathcal{S}, \lambda} - T_{\mu, \lambda}\|_\infty
    =\max_{\lambda \neq 0} \|T_{\nu_\mathcal{S}, \lambda}\|_\infty,
\end{gather}
where maximization is performed over all unique irreps appearing in the decomposition of $U^{t,t}$. In the simplest case, $d=2$, these are all $SU(2)$ representations with integer spin quantum number $s \le t$. For $d>2$, the irreps are labeled by the $d-1$-dimensional generalizations of a spin number (e.g. Young tableaux), and thus, more complicated conditions are required \cite{Dulian_2023, Dulian_2024, benkart1994, barut}. In either case, the dimensions of the $\pi_\lambda$ are $\mathcal{O}(t^{d(d-1)/2})$ and thus the norms $\|T_{\nu_\mathcal{S}, \lambda}\|_\infty$ can be computed efficiently.

\end{document}